\documentclass[aps,prx,reprint,superscriptaddress,amsmath,amssymb,floatfix,longbibliography]{revtex4-2}

\usepackage{graphicx}
\usepackage{dcolumn}
\usepackage{bm}
\usepackage[utf8]{inputenc}
\usepackage{physics}
\usepackage{mathtools}
\usepackage{booktabs}
\usepackage{algpseudocode}
\usepackage{xcolor}
\usepackage[most]{tcolorbox}
\usepackage[
  colorlinks=true,
  citecolor=red,
  linkcolor=blue,
  urlcolor=blue
]{hyperref}
\usepackage[normalem]{ulem}
\usepackage{placeins}

\newcounter{boxedalgorithm}
\renewcommand{\theboxedalgorithm}{\arabic{boxedalgorithm}}
\newenvironment{boxedalgorithm}[2]{%
  \refstepcounter{boxedalgorithm}%
  \begin{tcolorbox}[
    enhanced,
    colback=white,
    frame hidden,
    boxrule=0pt,
    borderline north={0.5pt}{0pt}{black},
    borderline south={0.5pt}{0pt}{black},
    left=2pt, right=2pt, top=6pt, bottom=6pt,
    before skip=10pt, after skip=10pt
  ]
  \noindent\textbf{Algorithm~\theboxedalgorithm. #1}\label{#2}\par
  \smallskip\hrule\smallskip
}{%
  \end{tcolorbox}%
}

\begin{document}

\title{Learning a quantitative criterion for distinguishing chaos from noise}

\author{Jaesung Choi}
\affiliation{Center for Artificial Intelligence and Natural Sciences, Korea Institute for Advanced Study, Seoul 02455, Korea}
\author{Athokpam Langlen Chanu}
\affiliation{Asia Pacific Center for Theoretical Physics, Pohang, 37673, Republic of Korea}
\affiliation{Department of Physics, Pohang University of Science and Technology, Pohang 37673, Republic of Korea}
\author{Jong-Min Park}
\email[Contact author: ]{jongmin.park@apctp.org}
\affiliation{Asia Pacific Center for Theoretical Physics, Pohang, 37673, Republic of Korea}
\affiliation{Department of Physics, Pohang University of Science and Technology, Pohang 37673, Republic of Korea}

\date{\today}

\begin{abstract}
Distinguishing chaos from noise using time-series data is fundamentally challenging because both exhibit irregular fluctuations and share many statistical and dynamical characteristics. Existing methods face two key limitations: temporally correlated noise can yield spurious signatures of chaos, and analyses of scalar time series often require explicit choices of embedding parameters. Here, we propose a purely data-driven method for distinguishing chaos and noise based on a reservoir-computing framework with a cross-prediction scheme. In the proposed approach, the model is trained to predict the future change of a variable from its current value, thereby combining a short-term predictability test with a test of the smoothness of deterministic flows. The recurrent structure of reservoir computing enables effective prediction of high-dimensional chaotic dynamics even from scalar time series without explicit delay-coordinate reconstruction, while the cross-prediction framework strongly suppresses spurious predictive correlations arising from noise.
We apply the proposed method to diverse synthetic and empirical time series. Chaotic systems consistently yield strong correlations between the true and predicted future changes, whereas noise processes remain clearly separated in a low-correlation regime. The method also exhibits substantial robustness to practical limitations in empirical data, including measurement noise, limited data length, and increasing prediction lag. These results demonstrate that the squared Pearson correlation coefficient provides a simple quantitative criterion for distinguishing chaos from noise directly from observed time-series data.
\end{abstract}

\maketitle

\section{Introduction}
\label{sec:intro}

Unpredictability and irregularity of dynamics have long been attributed to the influence of unknown degrees of freedom, described as random noise until the discovery of chaos fundamentally altered this perspective~\cite{kantz2003nonlinear}.
This realization motivated efforts to determine whether the apparent unpredictability observed in empirical time series arises from deterministic chaos or stochastic noise.
Discrimination between chaos and noise using time-series data is intrinsically difficult because they exhibit remarkably similar features, including broadband spectra, rapidly decaying autocorrelations, irregular fluctuations, and limited long-term predictability~\cite{kennel1992method, rosso2007distinguishing, bradley2015nonlinear}.

A natural strategy is to examine whether a given time series exhibits key signatures of chaos such as positive Lyapunov exponents and finite correlation dimensions. Various methods have been developed to estimate quantities such as the correlation dimension~\cite{grassberger1983characterization,grassberger1983measuring}, Lyapunov exponents~\cite{wolf1985determining,wales1991calculating}, and the Kolmogorov entropy~\cite{grassberger1983estimation} directly from observed time-series data.
However, direct estimation of these quantities from empirical data can be misleading. Certain noise processes can yield apparently chaotic signatures~\cite{eckmann1992fundamental}, such as finite correlation dimensions~\cite{osborne1989finite, theiler1991some}, spurious positive Lyapunov exponents~\cite{dammig1993estimation, tanaka1996lyapunov,ikeguchi1997lyapunov}, and finite $K_2$ entropy~\cite{provenzale1991convergence}, when standard estimation procedures are blindly applied.
These difficulties have motivated the development of alternative approaches for distinguishing chaotic and stochastic dynamics.

Such alternative approaches include comparisons of the predictive performance of linear and nonlinear Volterra--Wiener models~\cite{barahona1996detection};
visibility-graph--based characterizations~\cite{lacasa2010description,luque2011feigenbaum,luque2012analytical,ravetti2014distinguishing};
the noise titration method~\cite{poon2001titration}; sequential cycle-correlation analysis~\cite{zhang2006detecting,zhang2006complex}; machine-learning approaches~\cite{wolpert1990detecting,borges2019learning,boaretto2021discriminating,zanin2022can}; recurrence analysis~\cite{prado2022direct,flauzino2025quantifying}; and ordinal-pattern--based analyses~\cite{rosso2007distinguishing, olivares2012contrasting, zunino2012distinguishing, quintero2015numerical,zanin2021ordinal},
such as entropy--complexity planes~\cite{rosso2007distinguishing}, the causality Shannon--Fisher plane~\cite{olivares2012contrasting}, forbidden-pattern diagnostics~\cite{amigo2006order,amigo2007true,amigo2008combinatorial,carpi2010missing,rosso2012causality}, and the permutation spectrum test~\cite{kulp2014discriminating}.
Although these methods have provided valuable insights, their reliability can be limited~\cite{freitas2009failure,gao2012detecting}, because they often rely on empirical contrasts between specific classes of processes rather than on fundamental dynamical properties distinguishing chaos from noise.

Several approaches have attempted to exploit the fundamental distinction that chaotic dynamics are deterministic, whereas noise processes are stochastic.
One class of methods is based on short-term predictability. For example, nonlinear forecasting methods identify chaos by examining whether future values of a time series can be predicted from its current value over short prediction horizons using simplex prediction~\cite{sugihara1990nonlinear, tsonis1992nonlinear}.
Another class of approaches exploits the smoothness of deterministic flows in phase space. Methods such as the test of determinism examine this property by analyzing coarse-grained flow patterns reconstructed from observed trajectories~\cite{kaplan1992direct,wayland1993recognizing,kaplan1993coarse,salvino1994smoothness,ortega1998smoothness}.

However, these methods have important limitations.
In particular, nonlinear forecasting can produce non-negligible prediction accuracy for temporally correlated noise, even though the underlying process is stochastic~\cite{ikeguchi1997difference}.
Moreover, both methods become increasingly difficult to apply~\cite{jeong1999test} in high-dimensional systems.
In many practical situations only a scalar time series corresponding to a single observable is accessible.
The phase space must then first be reconstructed using delay-coordinate embedding~\cite{takens1981detecting}.
The reliability of the analysis then becomes sensitive to the choice of embedding parameters and inappropriate choices can lead to misleading conclusions.
These limitations motivate a purely data-driven approach that exploits fundamental dynamical differences between chaos and noise.

Meanwhile, reservoir computing has emerged as a powerful framework for predicting and characterizing chaotic dynamics~\cite{jaeger2004harnessing,pathak2017using,pathak2018model}. It has been successfully applied to the reconstruction of attractors, model inference, and stability analysis of chaotic systems~\cite{lu2018attractor,lu2017reservoir,margazoglou2023stability}. It has even been shown to accurately forecast chaotic trajectories well beyond the Lyapunov time horizon~\cite{pathak2018model, fan2020long, zhai2023emergence}. More recently, reservoir computing has achieved unsupervised recovery of chaotic attractors from time series corrupted by extreme levels of noise, a setting where both conventional and deep learning approaches fundamentally struggle~\cite{choi2025signal,choi2025unsupervised}. Yet, despite its intimate connection to chaotic dynamics, reservoir computing has received surprisingly little attention as a tool for discriminating between chaotic and stochastic time series.

We propose an effective prediction scheme and combine it with reservoir computing to address two major limitations of conventional approaches. First, these approaches typically require explicit delay-coordinate embedding when only a scalar time series is available. Second, conventional prediction tasks remain vulnerable to spurious predictability induced by temporally correlated stochastic processes.
Reservoir computing exploits its recurrent memory to infer effective dynamical representations directly from scalar observations without explicit embedding reconstruction.
The cross-prediction task, where the prediction target and the input variable are different, combines short-term predictability with the identification of deterministic flows, thereby suppressing spurious predictive correlations arising from noise.
We demonstrate that the resulting squared Pearson correlation coefficient provides a simple quantitative criterion that clearly separates chaotic and stochastic dynamics in both synthetic and empirical datasets.

The remainder of this article is organized as follows. Section~\ref{sec:framework} introduces the cross-prediction reservoir-computing framework and the resulting classification criterion, and illustrates its advantages over conventional value-to-value and difference-to-difference prediction schemes using representative examples. Section~\ref{sec:main} evaluates the proposed method using synthetic systems with known ground truth and a diverse collection of empirical time series. Section~\ref{sec:robust} examines its robustness to additive measurement noise, limited time-series length, and increasing prediction lag. Section~\ref{sec:summary} summarizes the main findings and discusses their broader implications. Section~\ref{sec:methods} describes the prediction schemes, reservoir construction, training, cross-validation, and hyperparameter-optimization procedures. The definitions and generation procedures for all synthetic and empirical datasets are provided in Appendix~\ref{app:datasets}, while the optimized hyperparameter sets are listed in Appendix~\ref{app:hp}.

\section{Classification framework}
\label{sec:framework}

\subsection{Cross-prediction reservoir computing model}

Our objective is to identify whether the observed series originates from underlying chaotic dynamics or from noise.
As an extreme case, we consider that only a scalar time series $(x_1, x_2, \dots, x_L)$ of length $L$ associated with a single degree of freedom is accessible. Although we focus on scalar time series, the framework can be straightforwardly extended to multivariate time series.

Our key strategy is to determine whether a reservoir-computing model can learn the dynamical relation between the current observation and a future change.
We define the first difference as $\Delta x_{t+1} \equiv x_{t+2} - x_{t+1}$ and propose a cross-prediction scheme in which the machine is trained to predict $\Delta x_{t+1}$ from $x_t$.

\begin{figure*}
\includegraphics[width=1.8\columnwidth]{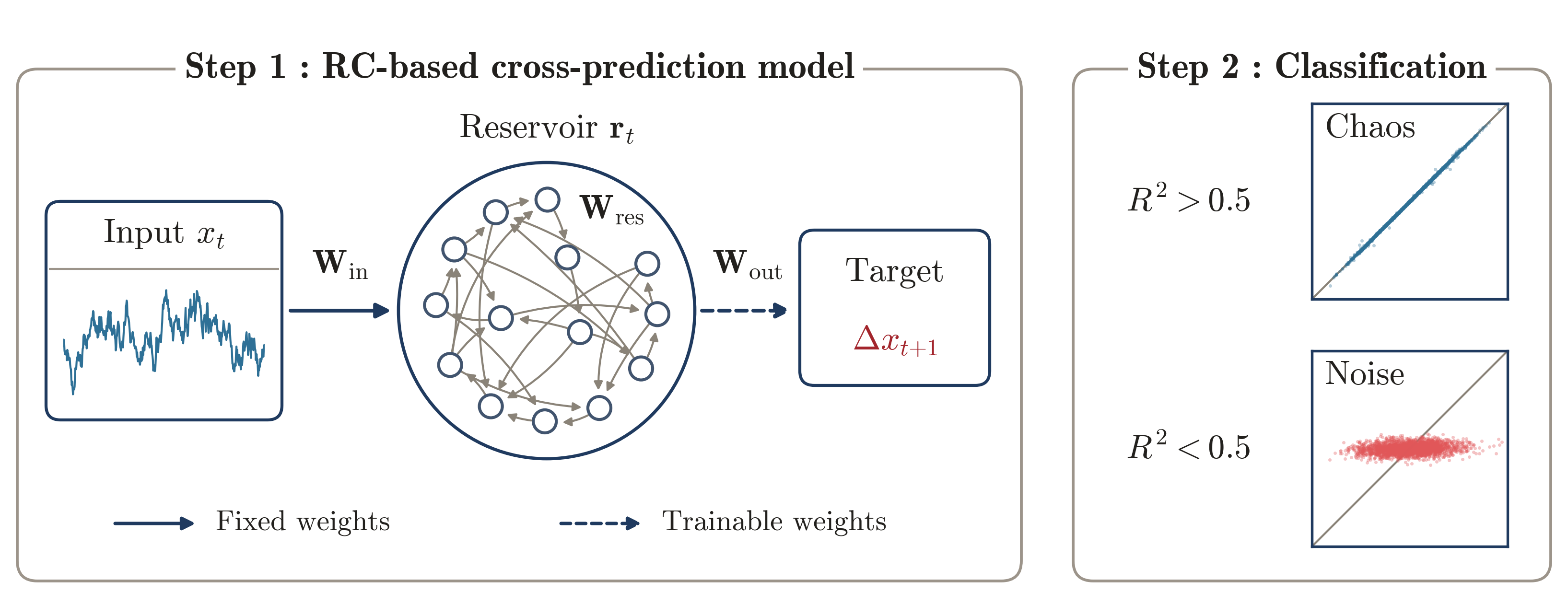}
\caption{\textbf{Schematic representation of the cross-prediction reservoir computing (RC) framework.}
Step~1 (left panel) corresponds to the RC-based prediction model, and Step~2 (right panel) corresponds to the classification criterion.
The reservoir receives a scalar input through fixed input weights $\mathbf{W}_{\mathrm{in}}$ and evolves through a fixed recurrent weight matrix $\mathbf{W}_{\mathrm{res}}$. Only the output weights $\mathbf{W}_{\mathrm{out}}$ are trained via ridge regression to predict the target.}
\label{fig:rc_scheme}
\end{figure*}

The overall framework, illustrated in Fig.~\ref{fig:rc_scheme}, consists of two steps. The first is training an RC model using the cross-prediction task and the second is classifying the time series according to the resulting prediction accuracy. The core definitions are given below, while the detailed training and hyperparameter-optimization procedures are described in the Methods (Sec.~\ref{sec:methods}).

Reservoir computing (RC) is a recurrent neural-network framework particularly suitable for reconstructing hidden dynamical information from scalar time series because of its intrinsic memory.
In a reservoir-computing model, only the linear readout layer is trained, while the recurrent reservoir and input weights are randomly initialized and kept fixed throughout training~\cite{jaeger2004harnessing}. This architectural choice is central to the discriminative power of the proposed method. The fixed recurrent reservoir and linear readout impose a restricted inductive bias, which reduces the tendency to memorize stochastic fluctuations compared with fully trainable recurrent architectures. As a result, a high held-out $R^2$ indicates learnable temporal structure consistent with an underlying deterministic rule under the proposed cross-prediction test.

We employ an echo state network (ESN) as the specific RC implementation. The reservoir state $\mathbf{r}_t \in \mathbb{R}^{n}$ with the number of reservoir nodes $n$ evolves according to the leaky-integrator update rule,
\begin{equation}
\mathbf{r}_{t} = (1 - a_{\mathrm{leak}})\,\mathbf{r}_{t-1}
+ a_{\mathrm{leak}}\tanh\!\left(\mathbf{W}_{\mathrm{res}}\,\mathbf{r}_{t-1}
+ \mathbf{W}_{\mathrm{in}}\,u_t\right),
\label{eq:reservoir}
\end{equation}
where $a_{\mathrm{leak}} \in (0,1]$ is the leak rate, $u_t$ is the scalar input at time $t$, and $\tanh(\cdot)$ is applied element-wise. Here $\mathbf{W}_{\mathrm{res}} \in \mathbb{R}^{n \times n}$ is a fixed recurrent weight matrix and $\mathbf{W}_{\mathrm{in}} \in \mathbb{R}^{n \times 1}$ is a fixed input-weight matrix; both are randomly initialized and are not updated during training.
Through this recurrent evolution, the reservoir naturally retains information from previous inputs, enabling it to construct an effective representation of the hidden dynamics without explicit phase-space reconstruction.

The reservoir state is then mapped to the predicted target through
a linear readout,
\begin{equation}
\hat y_t
=
\mathbf{W}_{\mathrm{out}}
\mathbf{X}_t ,
\end{equation}
where
$\mathbf{X}_t=[\mathbf r_t;1]$
denotes the extended reservoir state including the bias term.
Here $\hat y_t$ is the prediction of the target $y_t$.
The output weight matrix $\mathbf W_{\mathrm{out}}$ is
trained using the training dataset so that the error between the predicted output $\hat y_t$ and the target $y_t$ is minimized.
Details of the training process, including the construction of the reservoir and the roles of the leak rate $a_{\mathrm{leak}}$ and the spectral radius $\rho$, are provided in the Methods (Sec.~\ref{sec:methods}).

Although reservoir computing alleviates the need for explicit delay-coordinate embedding by exploiting its recurrent memory, conventional prediction tasks remain vulnerable to spurious predictability arising from strongly correlated stochastic processes.
Conventional short-term prediction tasks use the same observable for the input and target, for example
$u_t=x_t$ with $y_t = x_{t+1}$ or $u_t=\Delta x_t$ with $y_t = \Delta x_{t+1}$.
In contrast, our cross-prediction scheme uses
$u_t=x_t$ and $y_t=\Delta x_{t+1}$.
We refer to this task as cross-prediction because the input and target are different observables.
Successful cross-prediction implicitly combines two requirements: the predictability of the future observation $x_{t+1}$ from the available history and the learnability of the dynamical rule relating $x_{t+1}$ to its subsequent change $\Delta x_{t+1}$.
We also note that prior to training, the time series was centered and rescaled such that $\langle x_t \rangle =0$ and $\max |x_t| = 1$.
We use a five-fold cross-validation scheme (CV5), in which the time-ordered data are divided into five contiguous segments; each segment serves in turn as the validation set, while the remaining four are used for training.

We quantify the performance of the cross-prediction RC scheme using the squared Pearson correlation coefficient $R^2$ between the true target sequence $\{y_t\}$ and the predicted sequence $\{\hat{y}_t\}$, defined as
\begin{equation}
R^2 = \left(
\frac{\displaystyle \sum_t (y_t - \bar{y})(\hat{y}_t - \bar{\hat{y}})}
{\sqrt{\displaystyle \sum_t (y_t - \bar{y})^2 \displaystyle \sum_t (\hat{y}_t - \bar{\hat{y}})^2}}
\right)^{\!2},
\label{eq:r2}
\end{equation}
where $\bar{y}$ and $\bar{\hat{y}}$ are the averages of the target and the RC-predicted values, respectively.
$R^2$ is computed on each of the five validation folds and then averaged.
This definition takes values in $[0, 1]$ by construction.
Throughout this work, we operationally classify time series with $R^2 > 0.5$ as chaos and those with $R^2 < 0.5$ as noise.

\subsection{Results for representative examples}
\label{sec:rep}

\begin{figure}
\includegraphics[width=0.75\columnwidth]{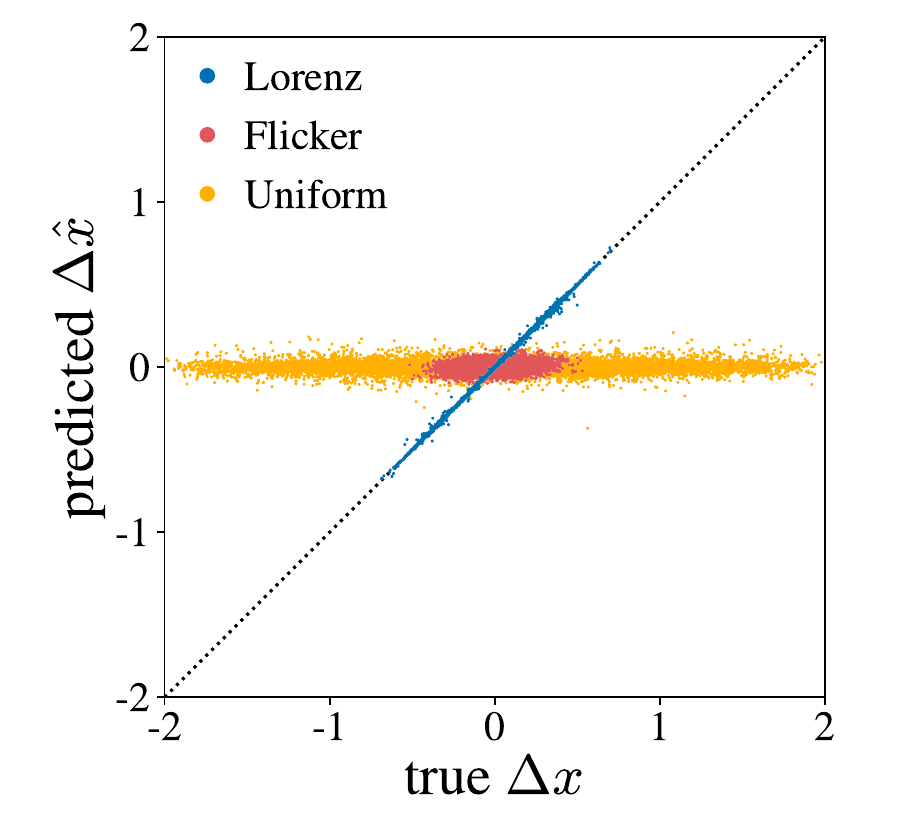}
\caption{
Scatter plot of the true and predicted values obtained using the proposed method. Blue, magenta, and yellow circles correspond to the Lorenz system, flicker noise, and uniform noise, respectively.
}
\label{fig:cor_scatter_main}
\end{figure}

To illustrate the effectiveness of the proposed cross-prediction scheme, we first apply it to three representative processes: (i) the chaotic Lorenz system, (ii) flicker noise with strong temporal correlations, and (iii) independent and identically distributed uniform noise.
Figure~\ref{fig:cor_scatter_main} presents the correlation between the true and predicted values obtained using the cross-prediction scheme.
A strong correlation is observed for the Lorenz system, indicating successful prediction even though the underlying system is three-dimensional and no explicit embedding-space reconstruction is performed.
In contrast, neither of the noise processes yields significant predictive correlations, enabling clear discrimination from chaos.

\begin{figure}
\includegraphics[width=\columnwidth]{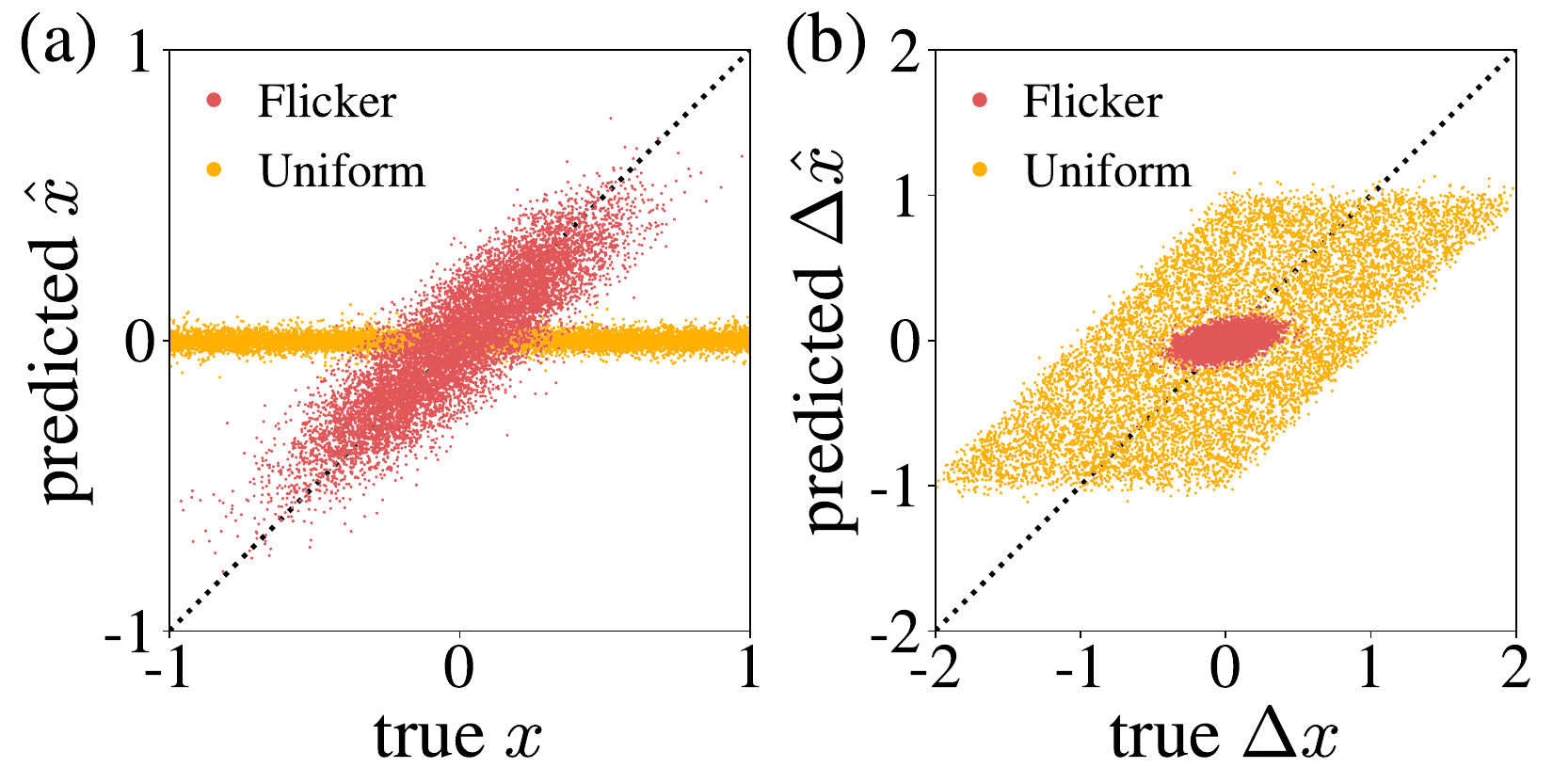}
\caption{
Scatter plots obtained using the conventional (a) value-to-value and (b) difference-to-difference prediction schemes. Magenta and yellow circles correspond to flicker noise and uniform noise, respectively.
}
\label{fig:cor_scatter_conventional}
\end{figure}

This result is nontrivial because conventional predictability-based approaches can produce spurious predictability for noise.
To demonstrate the limitations of conventional approaches, 
Fig.~\ref{fig:cor_scatter_conventional} shows the corresponding results for conventional value-to-value and difference-to-difference prediction schemes.
Flicker noise yields spurious predictability in the value-to-value scheme, whereas differencing suppresses this effect~\cite{ikeguchi1997difference}. In contrast, taking the first-difference induces spurious correlations for uniform white noise.
These results show that neither conventional scheme universally eliminates predictive correlations arising from noise.
To overcome these spurious-correlation problems, previous studies have relied on analyzing how prediction accuracy changes with increasing prediction horizons.
This highlights the main advantage of the proposed cross-prediction scheme, i.e., it successfully suppresses such correlation-induced predictability, enabling reliable discrimination between chaos and noise even in one-step prediction.

The intuitive explanation for the success of the proposed method is as follows. First, reservoir computing possesses a recurrent structure, in which the internal states evolve depending on both the current input and previous states. Thus, although no explicit delay-coordinate vector is supplied as input, past observations continue to influence the prediction through the recurrent reservoir state. In this sense, the reservoir implicitly learns an effective delay-coordinate representation optimized for the prediction task without explicitly specifying an embedding dimension.

Second, the proposed cross-prediction scheme imposes a stricter constraint by combining short-term predictability with a deterministic-flow test. Predicting the next difference from the current value requires the model not only to anticipate future evolution, but also to infer the underlying dynamical flow governing the local change of the system.
Noise processes generally do not satisfy both conditions simultaneously.
As a result, spurious predictive correlations arising from noise are strongly suppressed.

\section{Main results and discussion}
\label{sec:main}

\subsection{Synthetic datasets with known ground truth}
\label{sec:synth}

\begin{figure}
\includegraphics[width=\columnwidth]{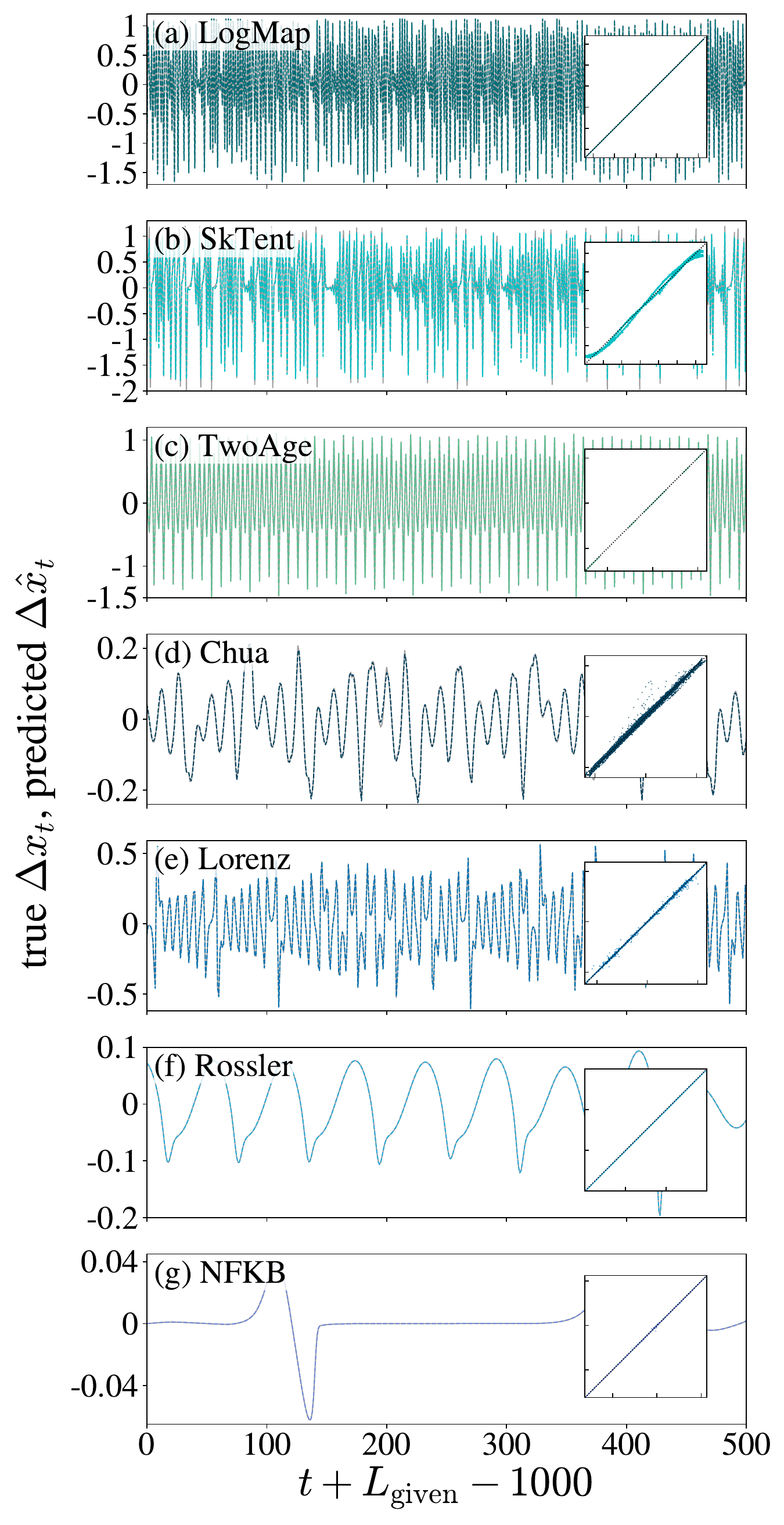}
\caption{
Comparison of the first-difference time series $\Delta x_t$ obtained from numerical simulations (gray) with the corresponding predictions $\Delta \hat{x}_t$ (colored) for the following synthetic chaotic systems: (a) logistic map, (b) skew tent map, (c) two-age-classes model, (d) Chua's circuit, (e) Lorenz system, (f) R\"ossler system, and (g) NF-$\kappa$B model. The main panels show 500-sample segments. The insets show scatter plots of $\Delta \hat{x}_t$ versus $\Delta x_t$, with the predicted and true values on the vertical and horizontal axes, respectively. Both axes in each inset span the same range as the vertical axis of the corresponding main panel.
}
\label{fig:pred_synth_chaos}
\end{figure}

\begin{figure}
\includegraphics[width=\columnwidth]{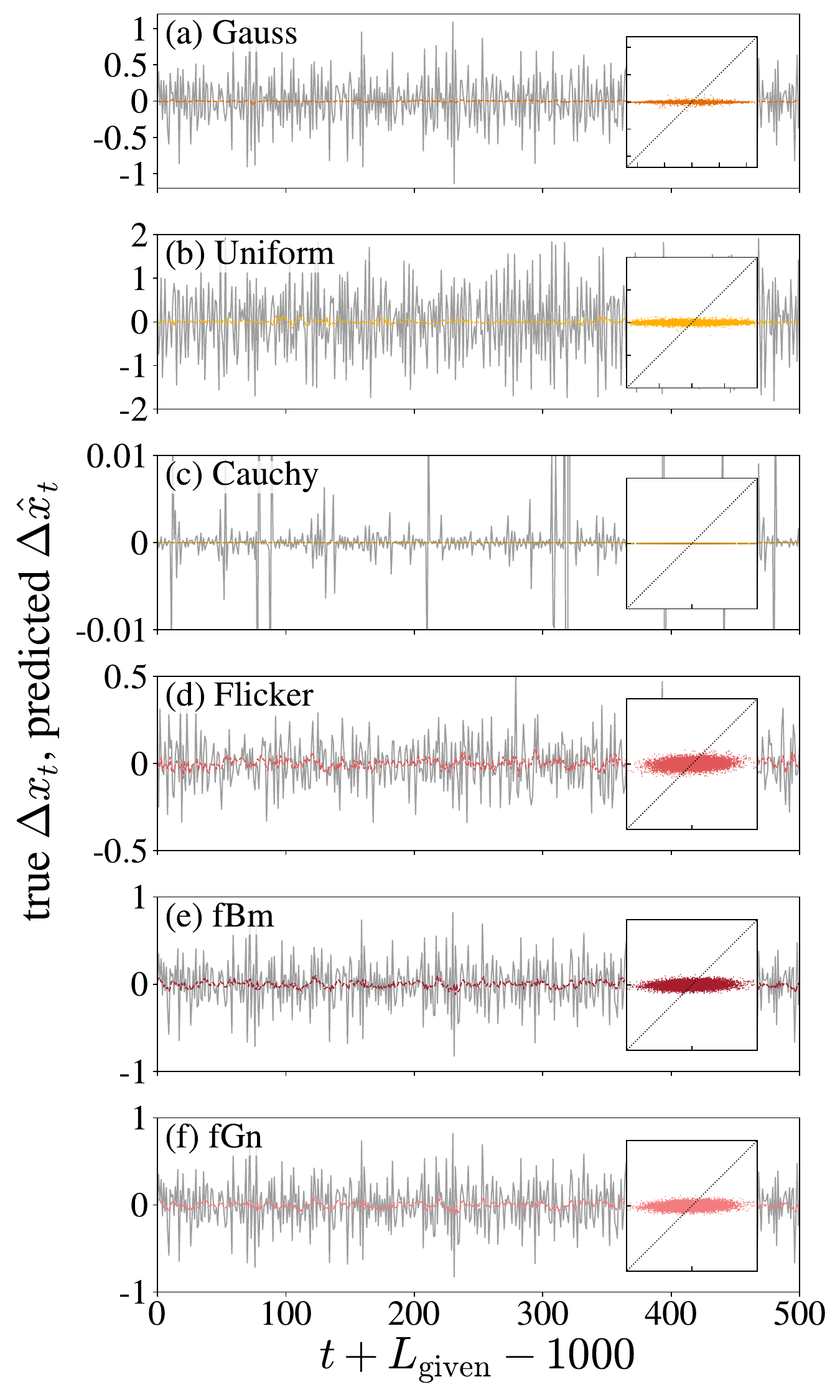}
\caption{
Comparison of $\Delta x_t$ (gray) with the corresponding predictions $\Delta \hat{x}_t$ (colored) for the following synthetic noise processes: (a) Gaussian noise, (b) uniform noise, (c) Cauchy noise, (d) flicker noise, (e) fractional Brownian motion, and (f) fractional Gaussian noise.
Plotting conventions as in Fig.~\ref{fig:pred_synth_chaos}.}
\label{fig:pred_synth_stoch}
\end{figure}

\begin{figure}
\includegraphics[width=0.9\columnwidth]{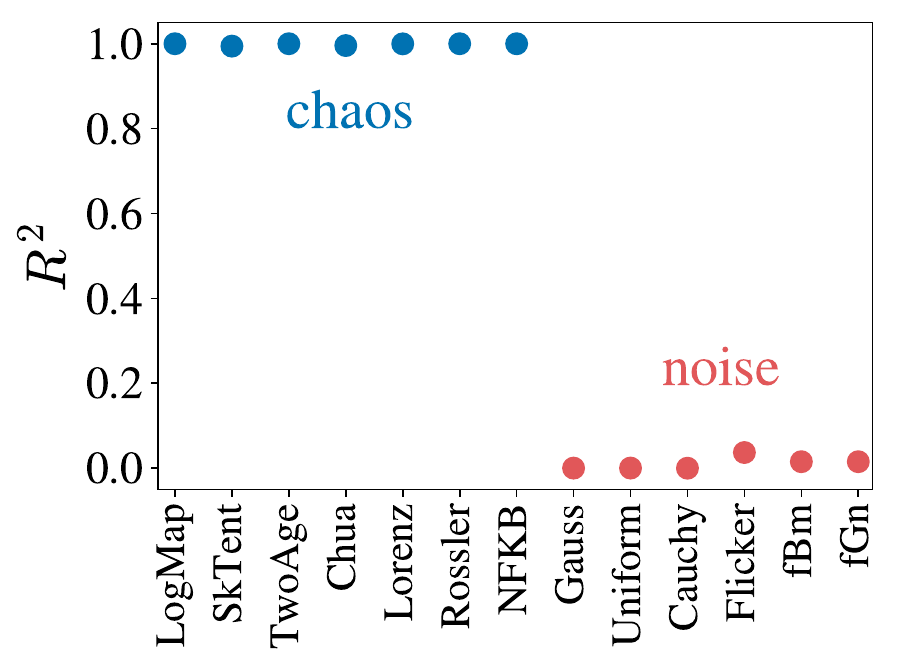}
\caption{
$R^2$ values for various simulated processes. Blue and red circles denote chaotic and noise processes, respectively.
}
\label{fig:R2_model}
\end{figure}

We further test the proposed method using synthetic time series generated from diverse chaotic systems and noise processes.
We first consider seven synthetic chaotic systems: the logistic map (LogMap), skew tent map (SkTent), two-dimensional density-dependent two-age-classes model (TwoAge), three-dimensional Chua's circuit (Chua), three-dimensional Lorenz system (Lorenz), three-dimensional R\"ossler system (Rossler), and a five-dimensional extended NF-$\kappa$B gene regulatory network model (NFKB).
These examples include both discrete-time chaotic maps and continuous-time chaotic systems. For all examples, we use only a single scalar time series associated with a single component.
Figure~\ref{fig:pred_synth_chaos} shows 500-sample segments of the true and predicted first-difference series, together with scatter plots computed from the full datasets. The prediction curves closely follow the true series, yielding near-perfect correlations.

We next consider six synthetic noise processes:
independent and identically distributed (i.i.d.) Gaussian noise (Gauss), i.i.d. uniform noise (Uniform), i.i.d. Cauchy noise (Cauchy), flicker noise (Flicker), fractional Brownian motion (fBm), and fractional Gaussian noise (fGn).
These processes include both temporally uncorrelated and correlated noise, as well as noise with infinite variance.
The resulting trajectories and correlation plots are shown in Fig.~\ref{fig:pred_synth_stoch}.
We observe that, unlike the chaotic systems, the prediction curves stay near zero and cannot capture the fluctuations in the true time series.
Thus, in the correlation plot, the data are vertically scattered around the origin, indicating essentially no correlation.

The resulting $R^2$ values for the chaotic systems and noise processes are presented in Fig.~\ref{fig:R2_model}.
The results clearly show that, for chaotic systems, $R^2$ is close to unity, whereas for noise, $R^2$ remains below approximately $0.1$.
Two examples are particularly noteworthy.
TwoAge is characterized by fragmented strange attractors, which are known to cause the failure of conventional nonlinear forecasting methods~\cite{cazelles1992predictable}.
Nevertheless, the proposed method clearly distinguishes TwoAge from stochastic noise.
Furthermore, despite its high dimensionality, the proposed method also achieves strong predictive performance for the five-dimensional NF-$\kappa$B system.
These examples demonstrate that the proposed method remains effective for challenging chaotic systems.
This clear separation demonstrates the effectiveness of our method in distinguishing simulated systems with known ground truth.
For each process, detailed descriptions of the numerical simulations are provided in Appendix~\ref{app:datasets}.

\subsection{Empirical time series}
\label{sec:emp}

\begin{figure}
\includegraphics[width=\columnwidth]{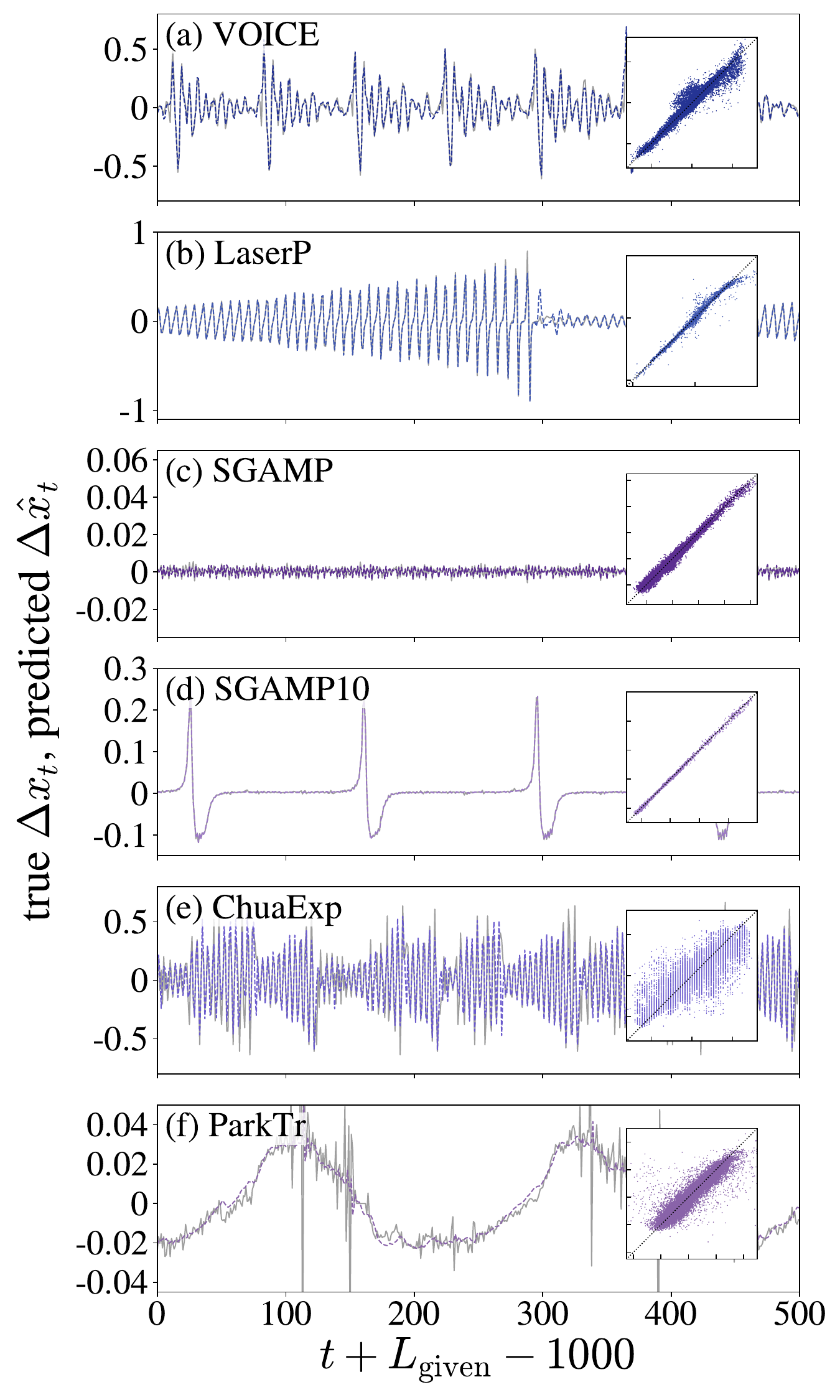}
\caption{
Comparison of $\Delta x_t$ (gray) with the corresponding predictions $\Delta \hat{x}_t$ (colored) for the following empirical time series: (a) voice recording, (b) laser intensity pulsation, (c) squid giant axon membrane potential (SGAMP), (d) SGAMP downsampled by a factor of 10, (e) experimental Chua's circuit, and (f) Parkinsonian tremor.
Plotting conventions as in Fig.~\ref{fig:pred_synth_chaos}.}
\label{fig:pred_emp_chaos}
\end{figure}

\begin{figure}
\includegraphics[width=\columnwidth]{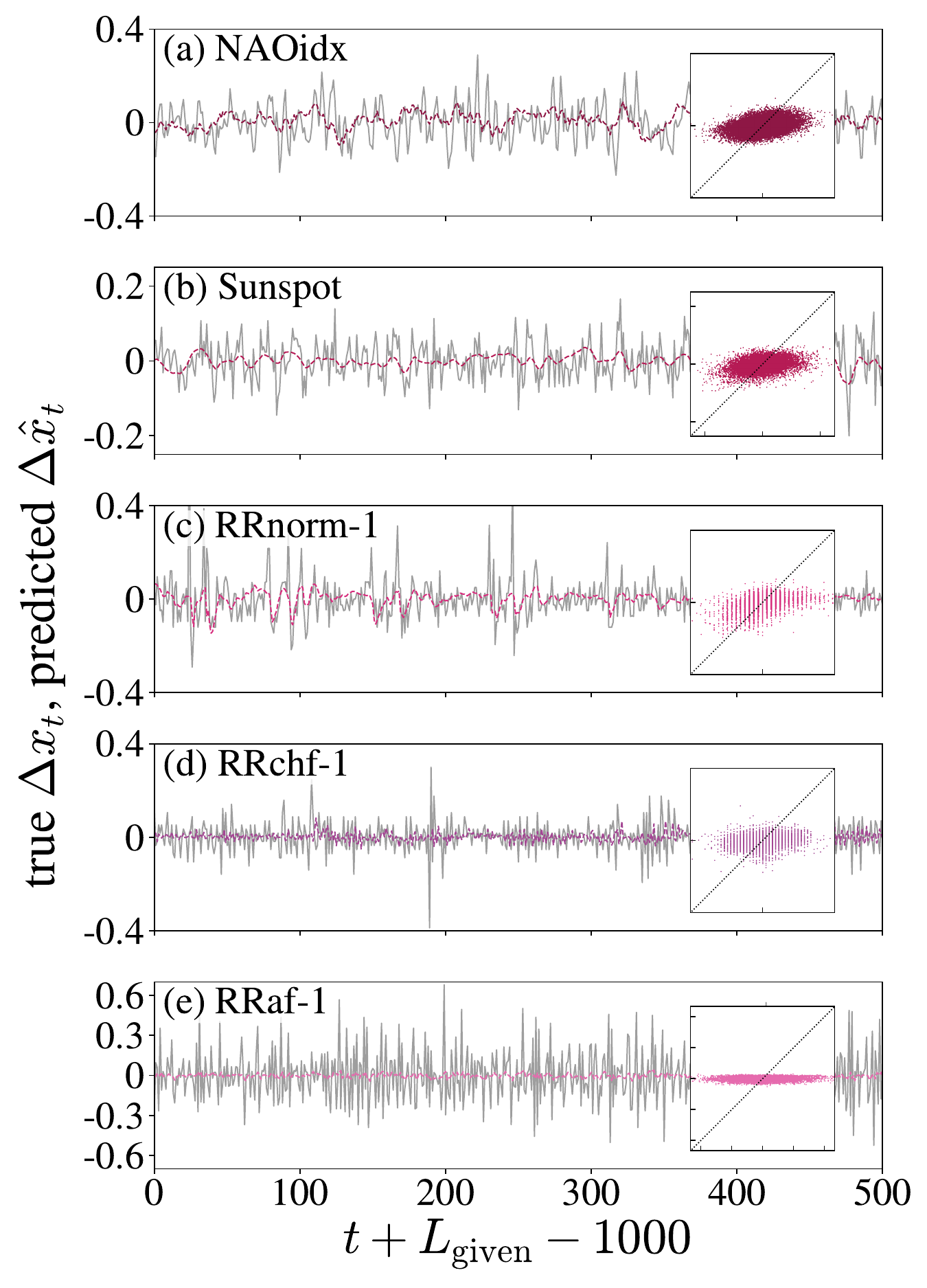}
\caption{
Comparison of $\Delta x_t$ (gray) with the corresponding predictions $\Delta \hat{x}_t$ (colored) for the following empirical time series: (a) North Atlantic Oscillation index, (b) sunspot number, and RR-interval time series from the first subject in each of three groups: (c) healthy subjects, (d) patients with congestive heart failure, and (e) patients with atrial fibrillation.
Plotting conventions as in Fig.~\ref{fig:pred_synth_chaos}.}
\label{fig:pred_emp_stoch}
\end{figure}

\begin{figure}
\includegraphics[width=0.9\columnwidth]{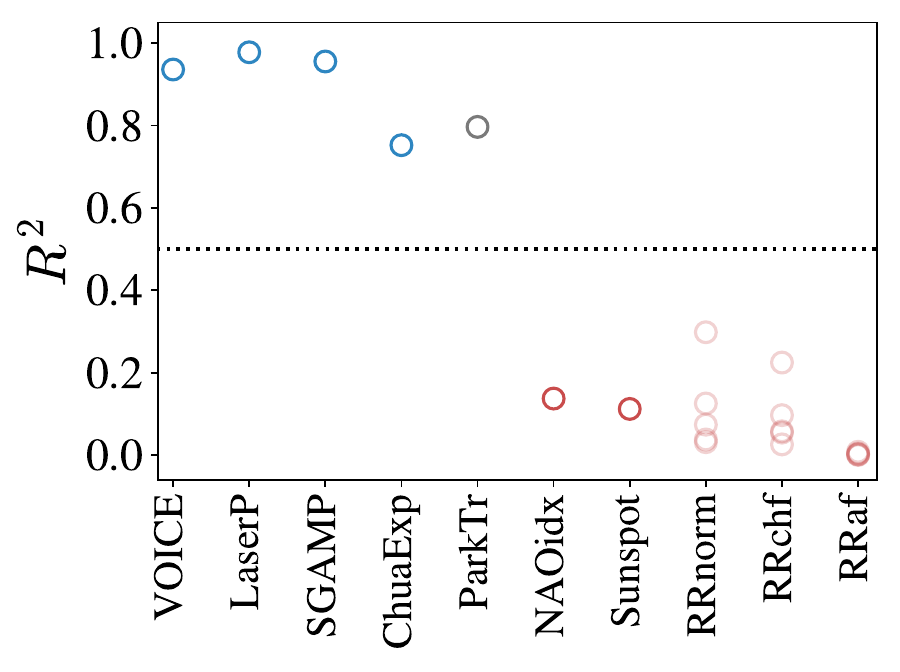}
\caption{
$R^2$ values for various empirical time series. Light blue and light red circles denote literature-based reference categories used for comparison; ParkTr is discussed separately because previous studies have reported conflicting classifications. For each type of RR interval time series, results from five independent datasets are plotted. The dotted horizontal line indicates the reference value $R^2 = 0.5$.
}
\label{fig:R2_real}
\end{figure}

We next turn to empirical time series to examine whether the proposed method remains effective beyond model-generated time series.
We first apply our method to the datasets obtained from voice recording (VOICE), laser intensity pulsation (LaserP), squid giant axon membrane potential (SGAMP), and an experimental Chua's circuit (ChuaExp), which have previously been identified as chaotic, together with Parkinsonian tremor (ParkTr) whose classification remains debated.
Figure~\ref{fig:pred_emp_chaos} compares the true and predicted first-difference series. The predictions agree closely with VOICE, LaserP, and SGAMP, although the correlations are generally lower than for the synthetic chaotic systems.

Lower correlations are observed for ChuaExp and ParkTr, likely for different reasons. For ChuaExp, the reduced performance is mainly attributable to coarse measurement resolution, as the analyzed voltage series is quantized in $0.2$ increments and contains only 69 distinct values. For ParkTr, the model captures much of the smooth temporal evolution but fails to reproduce several irregular peaks, suggesting mixed stochastic--deterministic dynamics with a substantial learnable deterministic component. This interpretation is consistent with the conflicting classifications reported in previous studies~\cite{gantert1992analyzing,sadeghirazlighi2012study, gao2002pathological,sarbaz2020exploring}.

SGAMP has a much finer temporal resolution and is approximately one order of magnitude longer than the other empirical time series. Consequently, a 1000-sample window shows only a small fraction of its overall dynamical structure. We therefore also analyze a modified version downsampled by a factor of 10, denoted SGAMP10. The insets in Fig.~\ref{fig:pred_emp_chaos}(c) and (d) show that SGAMP10 exhibits a stronger correlation between the true and predicted values than SGAMP. This result suggests that the dynamics appear more deterministic at a tenfold coarser temporal resolution, possibly because downsampling suppresses high-frequency noise.
To examine how this effect influences the results, we include both SGAMP and SGAMP10 in the subsequent robustness analyses.

We also apply our method to empirical time series previously identified as noise: the North Atlantic Oscillation index (NAOidx), sunspot number (Sunspot), and RR-interval recordings.
The RR-interval datasets comprise five recordings from different subjects in each of three groups involving healthy subjects (RRnorm), patients with congestive heart failure (RRchf), and patients with atrial fibrillation (RRaf).
The recordings are denoted by RRnorm-$n$, RRchf-$n$, and RRaf-$n$, where $n=1,\ldots,5$ indexes the subject within each group.
Figure~\ref{fig:pred_emp_stoch} shows NAOidx, Sunspot, and one representative recording from each RR group (RRnorm-1, RRchf-1, and RRaf-1).
Unlike the chaotic reference datasets, the predictions differ substantially from the observed data and fail to capture the pronounced fluctuations present in the observations.

Figure~\ref{fig:R2_real} presents $R^2$ values for empirical time series.
The obtained $R^2$ values demonstrate a consistent overall tendency, albeit with a less clear separation than in the model-generated time series, which is likely attributable to measurement noise, finite resolution, and other uncontrollable factors inherent in real data.
In particular, time series in the reference chaos category consistently yield high values of $R^2$, typically exceeding 0.6, whereas those in the reference noise category exhibit low $R^2$ values, generally below 0.4.
These observations indicate that the proposed method provides clear separation even in real-world data, with the reference chaos and noise categories occupying distinct high and low $R^2$ regimes separated around $R^2 \approx 0.5$.
Our criterion places ParkTr on the chaos side of the threshold. Its intermediate $R^2$ relative to the established chaotic datasets is consistent with mixed stochastic--deterministic dynamics containing a substantial learnable deterministic component.

\section{Robustness analyses}
\label{sec:robust}

\begin{figure}
\includegraphics[width=0.9\columnwidth]{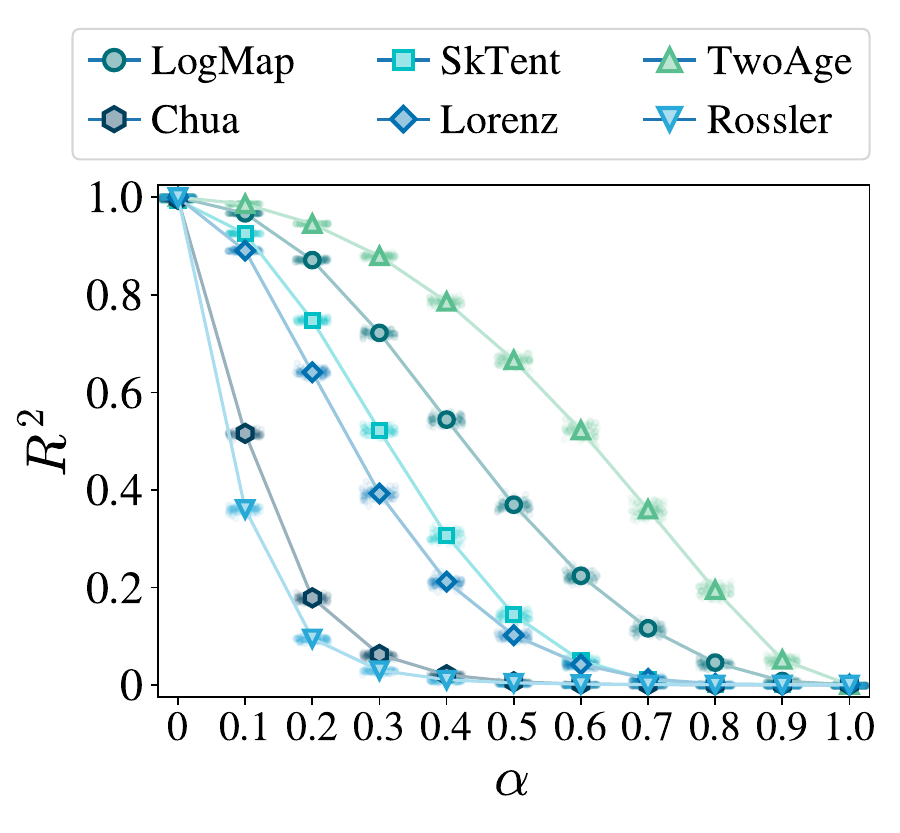}
\caption{$R^2$ as a function of the noise-mixing coefficient $\alpha$ for chaotic time series contaminated with Gaussian noise. Symbols represent the results for LogMap (circles), SkTent (squares), TwoAge (upward triangles), Chua (hexagons), Lorenz (diamonds), and R\"{o}ssler (downward triangles).}
\label{fig:noise_mixing}
\end{figure}

We now investigate the robustness of the proposed method to practical limitations in empirical data. We consider three factors involving the addition of white Gaussian noise, limitations in data length, and increases in prediction lag.

\subsection{Effect of additive noise}
\label{subsec:noise_mix}

Beyond the binary classification of pure chaos or pure noise, we further investigate how the proposed RC-based cross-prediction classifier responds when Gaussian noise is progressively added to chaotic time series, thereby interpolating between the two extremes.

Given a pure chaotic signal $x_t$, we construct a noise-contaminated signal via the convex-energy mixing rule
\begin{equation}
z_t = \sqrt{1 - \alpha^2}\,x_t + \alpha\,\eta_t,
\qquad \alpha \in [0, 1],
\label{eq:mixing}
\end{equation}
where $\alpha$ controls the noise fraction, and $\eta_t$ is Gaussian white noise with the standard deviation $\sigma = \mathrm{std}(x)$, ensuring that the original signal and bare noise have the same fluctuation scale. Since $(\sqrt{1-\alpha^2})^2 + \alpha^2 = 1$, the total signal energy is approximately preserved across all mixing levels. The coefficient is swept over $\alpha \in \{0, 0.1, 0.2, \dots, 1.0\}$ (11 levels). At $\alpha = 0$, the input is the original chaotic signal. At $\alpha = 1$, it reduces to pure Gaussian noise with the same variance.

We present the results of noise mixing on the chaotic signals in Fig.~\ref{fig:noise_mixing}.
It shows that $R^2$ decreases continuously as the noise level increases, although the rate of degradation varies considerably among the systems.
The discrete maps considered here generally exhibit a more gradual decrease than the continuous-time systems.
The trajectories shown in Fig.~\ref{fig:pred_synth_chaos} suggest that the sensitivity to noise may be related to both the magnitude and temporal structure of the one-step variations $\Delta{x}_t$.
Systems whose increments span a smaller range and vary more smoothly tend to lose predictive performance more rapidly.
Nevertheless, the differences among the curves cannot be fully explained by these simple features alone, suggesting that the response to noise also depends on the detailed dynamical structure of each system.
More importantly, $R^2$ does not decrease abruptly under weak noise contamination. 
Instead, the inferred classification remains unchanged over a finite range of noise levels and shifts to noise only when the added noise becomes sufficiently strong to substantially degrade the predictive information carried by the deterministic dynamics.
These results show that the proposed criterion tolerates a finite level of noise contamination while continuously reflecting the degradation of deterministic predictability.

\subsection{Effect of time series length}
\label{subsec:datlen}

In practice, the available length of a time series is often limited. To characterize the dependence of the proposed method on data length $L_{\mathrm{given}}$, we repeat the cross-prediction analysis for various values of $L_{\mathrm{given}} = 100, 200, 500, 1000, 2000, 5000, 10000$.

\begin{figure}
\includegraphics[width=\columnwidth]{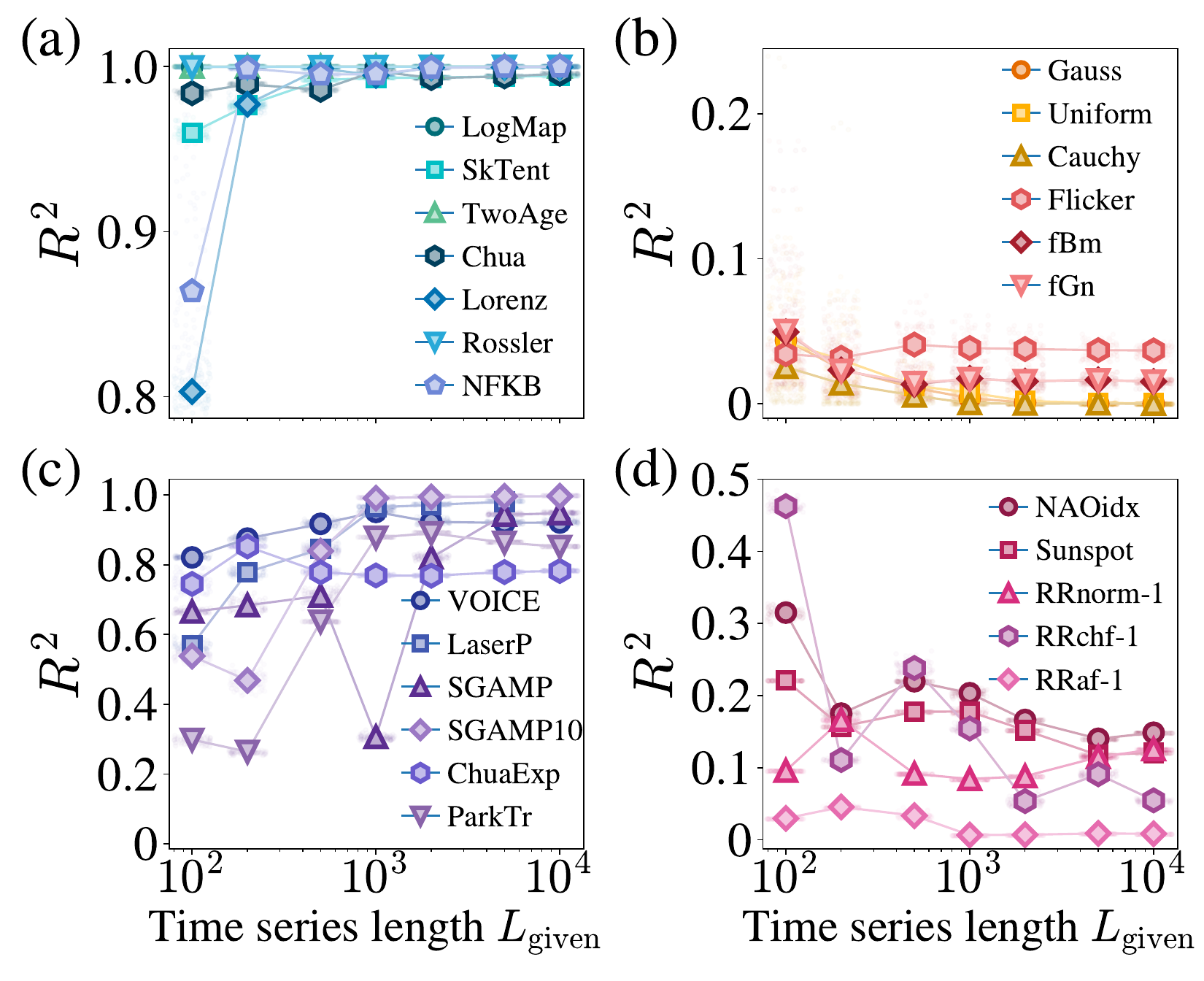}
\caption{$R^2$ as a function of time series length $L_{\mathrm{given}}$ for the examined datasets.
(a) Synthetic chaos. (b) Synthetic noise. (c) Empirical datasets in the chaos reference category. (d) Empirical datasets in the noise reference category. Faint dots represent individual realizations; solid lines connecting markers show the median across realizations.}
\label{fig:datlen}
\end{figure}

Figure~\ref{fig:datlen} shows $R^2$ as a function of $L_\mathrm{given}$ for all dataset categories. For synthetic chaos in panel (a), $R^2$ remains close to unity across nearly all time series lengths. Only at $L_\mathrm{given} = 100$ do a few systems start lower, around $0.8$, before recovering to near unity by $L_\mathrm{given} = 200$.
For synthetic noise in panel (b),  $R^2$ stays low ($<0.1$) across all lengths. A few processes show mildly elevated values at $L_\mathrm{given} = 100$ from finite-sample effects, which decrease toward zero as $L_\mathrm{given}$ increases.

\begin{figure}
\centering
\includegraphics[width=1\columnwidth]{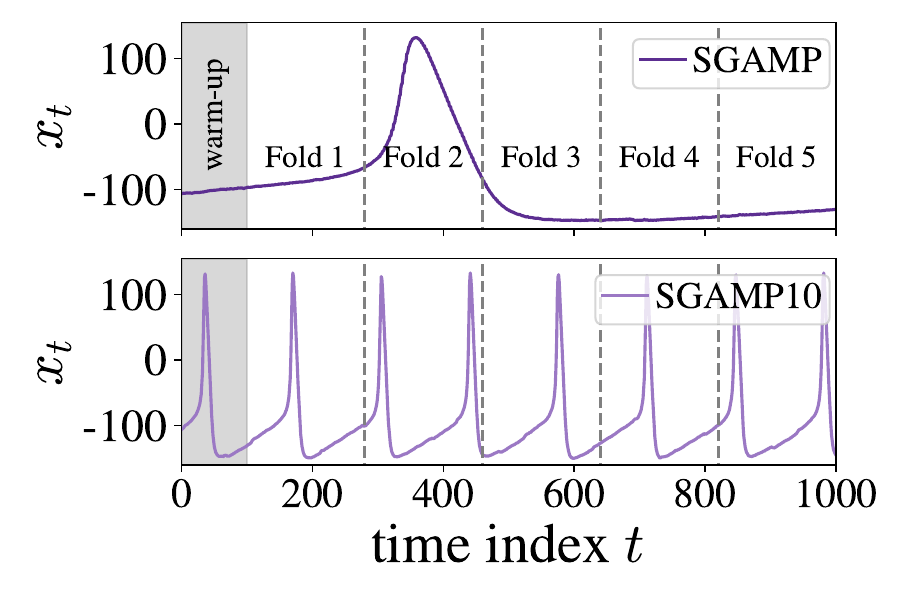}
\caption{Comparison of the first $1{,}000$ samples of SGAMP (top panel) and its $10{:}1$ downsampled version SGAMP10 (bottom panel), shown with the training setup at $L_\mathrm{given} = 1{,}000$. The shaded region is the first $100$ samples used for warm-up, and the dashed lines mark the five cross-validation fold boundaries.}
\label{fig:sgamp_comparison}
\end{figure}

For empirical datasets identified as chaos in panel (c), $R^2$ generally increases with $L_\mathrm{given}$ and stabilizes above the classification threshold. Most datasets exceed 0.8 at $L_\mathrm{given} = 10{,}000$. A notable case is SGAMP, whose $R^2$ shows a non-monotonic dip at $L_\mathrm{given} = 1{,}000$ before recovering at longer lengths.
To explain this, Fig.~\ref{fig:sgamp_comparison} shows a representative segment of SGAMP at this length, where the recording contains only a single action-potential spike. Within the five cross-validation folds, the spike falls in only one or two folds, while the remaining folds remain near a constant baseline. Most folds therefore have almost no dynamical variation to predict, and $R^2$ collapses for those folds. The dip thus reflects how the lone spike interacts with the fold split, not a genuine loss of predictability.
This effect is largely removed when the analyzed segment spans several spike cycles and thus better represents the underlying dynamics.
The bottom panel of Fig.~\ref{fig:sgamp_comparison} shows that SGAMP10 contains several spike cycles within the same 1000-sample window, which may account for the improved predictive performance shown in Fig.~\ref{fig:datlen}(c).
For SGAMP10, the same window contains a spike in every fold, and across lengths its $R^2$ rises to stabilize above 0.8 from $L_\mathrm{given} = 500$ onward.
Overall, most datasets consistently yield the same classification as the full-length time series across all tested lengths, even when only a few hundred samples are used. These results therefore demonstrate strong robustness to data length, provided that the sampling scale allows the observed window to capture the relevant dynamical variations.
For empirical datasets in the noise reference category in panel (d), $R^2$ remains below the threshold across all lengths, with moderate variability at $L_\textrm{given}=100$.

\begin{table}[t]
\centering
\caption{Classification consistency across time-series lengths. Entries give the number of datasets classified consistently with the reference class divided by the total number of datasets in that category, using $R^2 > 0.5$ for chaos and $R^2 < 0.5$ for noise. The final column summarizes the identical results obtained at $L_\mathrm{given}=2{,}000$, $5{,}000$, and $10{,}000$.}
\label{tab:success_rate}
\begin{ruledtabular}
\begin{tabular}{lccccc}
Data type & \multicolumn{5}{c}{$L_\mathrm{given}$} \\
\cline{2-6}
& $100$ & $200$ & $500$ & $1000$ & $\geq 2000$ \\
\hline
Synthetic chaos & $7/7$ & $7/7$ & $7/7$ & $7/7$ & $7/7$ \\
Synthetic noise & $6/6$ & $6/6$ & $6/6$ & $6/6$ & $6/6$ \\
\hline
Empirical chaos & $5/6$ & $4/6$ & $6/6$ & $5/6$ & $6/6$ \\
Empirical noise & $5/5$ & $5/5$ & $5/5$ & $5/5$ & $5/5$ \\
\end{tabular}
\end{ruledtabular}
\end{table}

To summarize classification reliability across time series lengths, Table~\ref{tab:success_rate} reports the number of datasets classified consistently with the reference class for each dataset group and time series length. For the synthetic systems, all seven chaotic datasets and all six noise datasets are classified consistently at every length. For the empirical signals, the five noise-reference datasets are also consistent with their reference class at every length. The only discrepancies occur in the empirical-chaos group. ParkTr falls on the noise side of the threshold at $L_\mathrm{given} = 100$ and $200$, SGAMP10 at $L_\mathrm{given} = 200$, and SGAMP at $L_\mathrm{given} = 1{,}000$.
Importantly, for all datasets examined, the classification remains unchanged and $R^2$ remains nearly constant for $L_\mathrm{given} \geq 2000$. These results demonstrate strong robustness to data length at sample sizes reasonably attainable in practice.

\subsection{Effect of prediction lag}
\label{subsec:lag}

We investigate the effects of the prediction horizon, denoted by $\ell$, by training the model to predict the first difference $\Delta x_{t+\ell}$ rather than the one-step-ahead target $\Delta x_{t+1}$.
Increasing $\ell$ enlarges the temporal separation between the input and target.
To characterize this sensitivity, we sweep $\ell \in \{1, 2, 3, 4\}$ and evaluate $R^2$ for all datasets.

\begin{figure}
\includegraphics[width=\columnwidth]{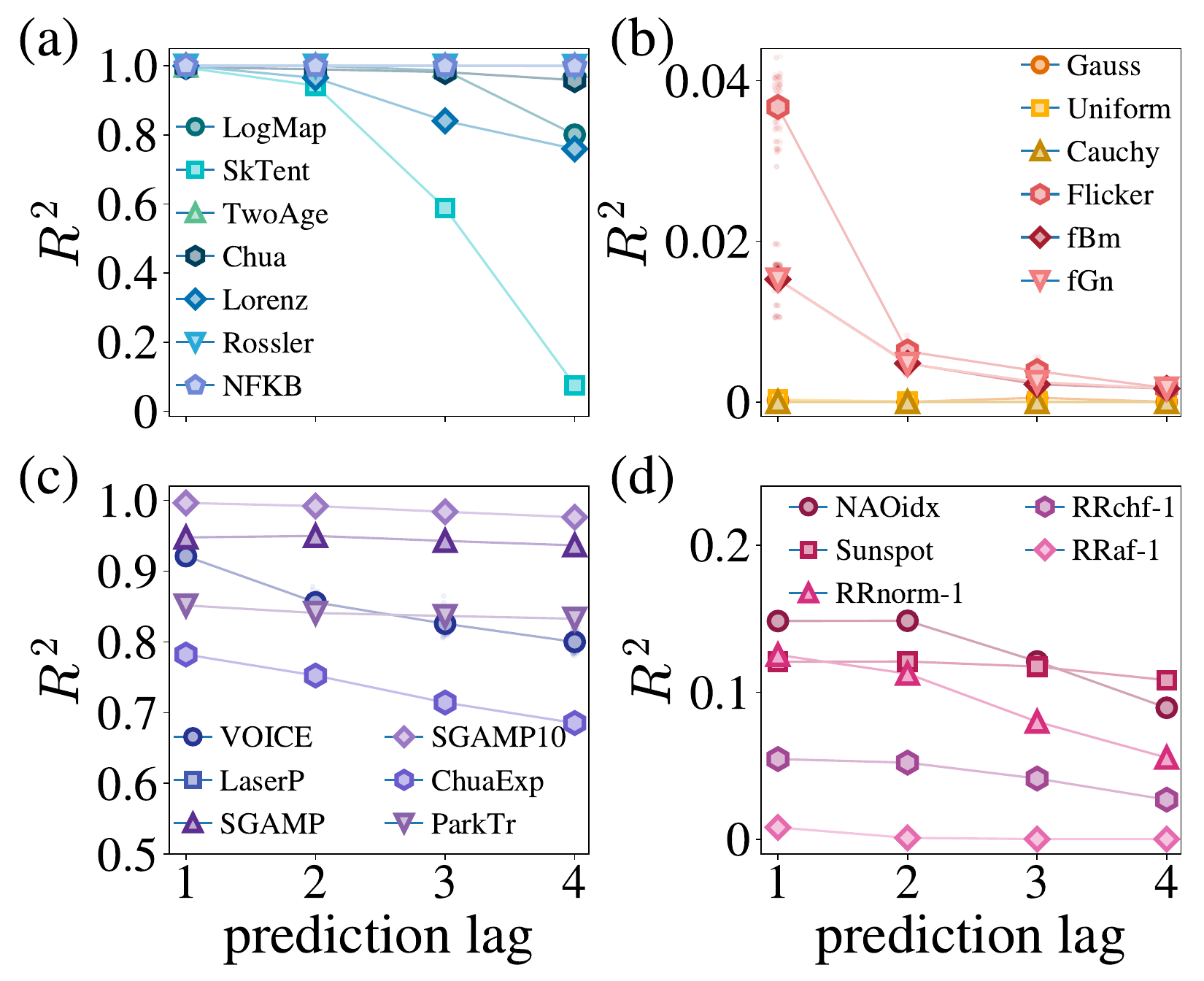}
\caption{$R^2$ as a function of prediction lag for the examined datasets. (a) Synthetic chaos. (b) Synthetic noise. (c) Empirical datasets in the chaos reference category. (d) Empirical datasets in the noise reference category. Faint dots show individual realizations; solid lines connecting markers show the median across realizations.}
\label{fig:lag}
\end{figure}

Figure~\ref{fig:lag} shows $R^2$ as a function of $\ell$ across all dataset categories.
It shows that $R^2$ decreases monotonically as the prediction lag $\ell$ increases. 
For both synthetic and empirical noise, $R^2$ remains below the classification threshold throughout the examined range, so the prediction lag does not affect their classification.
Among the chaotic systems, only SkTent exhibits a sufficiently rapid decrease for the classification to change, with $R^2$ falling below $0.5$ at $\ell=4$.
The particularly rapid decrease observed for SkTent may reflect the combined effects of its relatively large Lyapunov exponent and the increasing complexity of its iterated dynamics.
Except for SkTent, the other chaotic processes, both synthetic and empirical, exhibit only modest decreases over the examined lag range.
These results show that the cross-prediction scheme remains effective at discriminating between chaos and noise even as the prediction horizon increases over the range examined.

\section{Conclusion}
\label{sec:summary}

The present work addresses two intrinsic limitations of conventional nonlinear forecasting for distinguishing deterministic chaos from stochastic noise. The first is its reliance on delay-coordinate embedding, whose performance depends sensitively on the choice of embedding parameters.
The second is that conventional prediction tests can yield high predictive performance for stochastic processes with strong temporal correlations, making deterministic and stochastic dynamics difficult to distinguish.
By combining reservoir computing with a cross-prediction task, the proposed framework overcomes both limitations, enabling purely data-driven discrimination between deterministic chaos and stochastic noise from scalar time series.

The proposed framework was validated using a broad range of synthetic and empirical time series.
It correctly identified chaotic dynamics, including in systems for which conventional nonlinear forecasting can fail, while remaining effective for higher-dimensional chaotic systems.
Furthermore, it maintained strong performance for experimental datasets and provided consistent conclusions for datasets whose dynamical nature has remained controversial.
Taken together, these results demonstrate that the proposed framework provides a practical and broadly applicable approach for prediction-based chaos--noise discrimination across a wide range of real-world time-series datasets.

Beyond binary chaos--noise classification, the cross-prediction score provides a continuous measure of how strongly the observed dynamics support a deterministic description.
Tracking $R^2$ while varying the observed degrees of freedom or the temporal scale may therefore reveal crossovers between regimes in which deterministic and stochastic descriptions are more appropriate.
This perspective may be particularly useful when unresolved degrees of freedom or changes in observation time scale alter the effective description of a system.

Although the present framework was developed using reservoir computing, the underlying cross-prediction strategy may also be implemented using other recurrent neural network architectures. We adopted reservoir computing because only the readout layer is trained, allowing recurrent dynamics to be exploited without optimizing a large number of parameters. Whether more expressive trainable recurrent architectures can further improve the proposed framework, or instead become more susceptible to overfitting to noise in empirical datasets, remains an interesting direction for future investigation.

\section{Methods}
\label{sec:methods}

\subsection{Prediction schemes}
\label{subsec:schemes}

Prior to training, each scalar series is centered and rescaled as
\begin{equation}
x_t \leftarrow x_t - \langle x \rangle,
\qquad
x_t \leftarrow \frac{x_t}{\max_{t'} |x_{t'}|},
\label{eq:preprocess}
\end{equation}
where $\langle x \rangle$ is the time average, so that the processed series satisfies $\langle x_t \rangle = 0$ and $\max_t |x_t| = 1$.

Reservoir computing-based forecasting can be implemented with different choices of input and target. We consider three prediction schemes, summarized in Table~\ref{tab:schemes}, and adopt the value-to-difference scheme as the main prediction scheme.

\begin{table}[h]
\centering
\caption{Summary of the three prediction schemes.}
\label{tab:schemes}
\begin{tabular}{lcc}
\hline
Prediction scheme & Input ($u_t$) & Target ($y_t$) \\
\hline
Value-to-value           & $x_t$        & $x_{t+\ell}$        \\
Difference-to-difference & $\Delta x_t$ & $\Delta x_{t+\ell}$ \\
Value-to-difference      & $x_t$        & $\Delta x_{t+\ell}$ \\
\hline
\end{tabular}
\end{table}

In each scheme the reservoir is driven sequentially by an input series, and the linear readout maps the reservoir state to a target. Because the reservoir state evolves recurrently through Eq.~\eqref{eq:reservoir}, the prediction at each step reflects the input history rather than the instantaneous input alone.

\subsection{Reservoir construction and training}
\label{subsec:training}

The reservoir weight matrix $\mathbf{W}_{\mathrm{res}} \in \mathbb{R}^{n \times n}$ is constructed by drawing a dense Gaussian random matrix, applying a Bernoulli sparsity mask with connection probability $k$, and rescaling the result so that the spectral radius equals a target value $\rho$. The input weight matrix $\mathbf{W}_{\mathrm{in}} \in \mathbb{R}^{n \times 1}$ has entries drawn i.i.d.\ from $\mathcal{N}(0, \sigma_{\mathrm{in}}^2)$, where $\sigma_{\mathrm{in}}$ is the input scaling hyperparameter. Both $\mathbf{W}_{\mathrm{res}}$ and $\mathbf{W}_{\mathrm{in}}$ are fixed after initialization and are not updated during training.

The leak rate $a_{\mathrm{leak}}$ controls the temporal memory of the reservoir. A value close to $1$ makes the reservoir respond rapidly to new inputs, while a value close to $0$ produces slow, integrative dynamics. The spectral radius $\rho$ governs the stability of the reservoir; values below 1 are commonly used to promote the echo state property, while values modestly above 1 can enhance the reservoir's sensitivity to initial conditions, which is beneficial for chaotic time series.

We construct the matrix $\mathbf{X} \in \mathbb{R}^{(n+1) \times L_{\mathrm{train}}}$ by stacking the extended state vectors $\mathbf{X}_t = [\mathbf{r}_t;\,1] \in \mathbb{R}^{n+1}$ as columns over the training time steps $t=1,\dots, L_{\mathrm{train}}$, where $L_{\mathrm{train}}$ denotes the length of the training time series.
The corresponding target values are arranged row-wise in the matrix $\mathbf{Y}$, and the output weights are obtained analytically from the ridge-regression expression as
\begin{equation}
\mathbf{W}_{\mathrm{out}}
= \mathbf{Y}\mathbf{X}^{\!\top}
\!\left(\mathbf{X}\mathbf{X}^{\!\top} + \beta\,\mathbf{I}\right)^{-1},
\label{eq:readout}
\end{equation}
which reduces to ordinary least squares when the regularization coefficient $\beta=0$.
Here, $\!\top$ denotes transpose, and $\mathbf{I}$ is the identity matrix.
Since the entire training procedure reduces to a single matrix inversion, the RC avoids the iterative gradient-based optimization required by deep learning architectures, eliminating issues such as vanishing gradients and sensitivity to learning rate schedules. More importantly for the present objective of distinguishing chaos from noise, the fixed reservoir and linear readout impose a restricted inductive bias. Even on the training set, the target can be fitted accurately only to the extent that it can be approximated by a linear combination of the nonlinear reservoir activations. Deep learning models with large parameter counts, by contrast, are susceptible to overfitting noisy time series data~\cite{langarica2023contrastive, liu2023unsupervised}, as their expressive nonlinear layers can memorize stochastic fluctuations rather than learning a generalizable deterministic structure.

The RC model has six hyperparameters to be tuned for each dataset, which are the reservoir size $n$, spectral radius $\rho$, input scaling $\sigma_{\mathrm{in}}$, leak rate $a_{\mathrm{leak}}$, ridge regularization coefficient $\beta$, and sparsity parameter $k$. The search bounds for each parameter are listed in Table~\ref{tab:bounds}.

\begin{table}[h]
\centering
\caption{Hyperparameter search bounds used in surrogate-based optimization.}
\label{tab:bounds}
\begin{tabular}{lll}
\hline
Variable & \ \ Meaning & Bounds \\
\hline
$n$ & reservoir size & $[50,\; 1000]$ \\
$\rho$ & spectral radius & $[0.1,\; 2.0]$ \\
$\sigma_{\mathrm{in}}$ & input scaling & $[0.01,\; 3.0]$ \\
$a_{\mathrm{leak}}$ & leak rate & $[0.01,\; 1]$ \\
$\beta$ & ridge regularization & $[10^{-10},\; 10^{-3}]$ \\
$k$ & sparsity parameter & $[0.01,\; 1]$ \\
\hline
\end{tabular}
\end{table}

We perform the optimization separately for each combination of dataset, prediction scheme, and forecast lag. The initial $L_{\mathrm{warm}} = 100$ samples are used only to evolve the reservoir state $\mathbf{r}_t$ and are discarded to eliminate the influence of the initial condition. For shorter time series with $L_{\mathrm{given}} < 1{,}000$, we instead use $L_{\mathrm{warm}} = 0.1 \times L_{\mathrm{given}}$ to preserve a sufficient fraction of samples for training. These samples are not included in the training of $\mathbf{W}_{\mathrm{out}}$. Before the fold split, we also omit terminal samples for which the corresponding prediction target is not defined; for example, in the cross-prediction scheme with $y_t = \Delta x_{t+\ell}$, the valid targets require $t \leq L_{\mathrm{given}} - \ell - 1$. We denote the remaining number of valid state--target pairs after the warm-up removal by $L_{\mathrm{eff}}$.

To assess the generalization performance of each hyperparameter configuration, we employ a sequential time-ordered five-fold cross-validation (CV5) scheme. For each reservoir realization, the reservoir is run once over the input sequence, and the resulting valid state--target pairs are partitioned into five contiguous folds $\mathrm{fold}_1, \ldots, \mathrm{fold}_5$ of equal size $L_{\mathrm{fold}} = \lfloor L_{\mathrm{eff}}/5\rfloor$. For each $v \in \{1, \ldots, 5\}$, $\mathrm{fold}_v$ serves as the validation set and the remaining four folds serve as the training set.
The readout weights $\mathbf{W}_{\mathrm{out}}$ are obtained analytically via ridge regression [Eq.~\eqref{eq:readout}], and the root-mean-square error (RMSE) on the validation fold is computed as
\begin{equation}
\mathrm{RMSE}_v = \sqrt{\frac{1}{L_{\mathrm{fold}}}
\sum_{t \in \mathrm{fold}_v} (y_t - \hat{y}_t)^2},
\label{eq:rmse}
\end{equation}
where $y_t$ and $\hat{y}_t$ are the true target and the predicted target, respectively.

For each hyperparameter configuration, we repeat the CV5 evaluation over 10 independent reservoir realizations. For the $j$th realization, we compute the mean validation RMSE across the five folds as
\begin{equation}
\overline{\mathrm{RMSE}}_{\mathrm{CV5}}^{(j)}
= \frac{1}{5} \sum_{v=1}^{5} \mathrm{RMSE}_v^{(j)} .
\label{eq:cv5rmse}
\end{equation}
The hyperparameter-optimization objective is defined as
\begin{equation}
\mathcal{L}_{\mathrm{opt}}
=
\frac{1}{5}
\sum_{j \in \mathcal{B}_5}
\overline{\mathrm{RMSE}}_{\mathrm{CV5}}^{(j)},
\label{eq:optloss}
\end{equation}
where $\mathcal{B}_5$ denotes the set of the five reservoir realizations with the smallest values of $\overline{\mathrm{RMSE}}_{\mathrm{CV5}}^{(j)}$ among the 10 realizations. This procedure reduces sensitivity to reservoir initialization while retaining hyperparameter configurations that perform well across independent realizations. We optimize the hyperparameters by minimizing $\mathcal{L}_{\mathrm{opt}}$ using surrogate-based global optimization as implemented in MATLAB's \texttt{surrogateopt} with 500 iterations.

After obtaining the optimized hyperparameter set, we report the final performance using the CV5 mean of the squared Pearson correlation coefficient $R^2$ (Sec.~\ref{sec:framework}), which is bounded in $[0,1]$ and hence comparable across datasets of different amplitudes. To further reduce sensitivity to reservoir initialization, we evaluate the optimized hyperparameter set across 200 independent reservoir realizations and report the $R^2$ distribution of the top 100 realizations ranked by $\overline{\mathrm{RMSE}}_{\mathrm{CV5}}$. Algorithm~\ref{alg:rc} summarizes the overall procedure of the cross-prediction-based reservoir computing.

\begin{widetext}
\begin{boxedalgorithm}{Cross-prediction reservoir computing for chaos--noise discrimination}{alg:rc}
\small
\begin{algorithmic}[1]
\Statex \textbf{Input:} Scalar time series $\{x_t\}_{t=1}^{L}$;\; prediction lag $\ell$;\; hyperparameter search space $\mathcal{H}$ over $(n, \rho, \sigma_{\mathrm{in}}, a_{\mathrm{leak}}, \beta, k)$
\Statex \textbf{Output:} $R^2$ score and classification label
\Statex

\State Construct targets $y_t \gets \Delta x_{t+\ell}$ for $t = 1, \ldots, L-\ell-1$

\For{each candidate $h \in \mathcal{H}$ in the surrogate-based search}
    \For{$j = 1$ \textbf{to} $10$}
        \State Initialize an independent reservoir realization with hyperparameters $h$
        \State Run the reservoir over the input sequence
        \State Discard the first $L_{\mathrm{warm}}$ valid state--target pairs; split the remaining pairs into 5 sequential folds
        \For{$v = 1$ \textbf{to} $5$}
            \State Form $\mathbf{X}_{\mathrm{tr}} = [\mathbf{r}_t;\,1]_{t \notin \mathrm{fold}_v}$,\; $\mathbf{Y}_{\mathrm{tr}} = [y_t]_{t \notin \mathrm{fold}_v}$
            \State $\mathbf{W}_{\mathrm{out}}^{(h,j,v)} \gets \mathbf{Y}_{\mathrm{tr}}\mathbf{X}_{\mathrm{tr}}^{\!\top}
            \left(\mathbf{X}_{\mathrm{tr}}\mathbf{X}_{\mathrm{tr}}^{\!\top} + \beta\,\mathbf{I}\right)^{-1}$
            \State Compute $\mathrm{RMSE}_v^{(j)}(h)$ on validation fold $v$
        \EndFor
        \State $\overline{\mathrm{RMSE}}_{\mathrm{CV5}}^{(j)}(h)
        \gets \tfrac{1}{5} \sum_{v=1}^{5} \mathrm{RMSE}_v^{(j)}(h)$
    \EndFor
    \State $\mathcal{L}_{\mathrm{opt}}(h) \gets$ average of the five smallest values among
    $\{\overline{\mathrm{RMSE}}_{\mathrm{CV5}}^{(j)}(h)\}_{j=1}^{10}$
\EndFor

\State $h^{\star} \gets \arg\min_{h}\,\mathcal{L}_{\mathrm{opt}}(h)$
\State Evaluate 200 independent reservoir realizations using $h^{\star}$
\State Report the distribution of CV5 mean $R^2$ over the top 100 realizations ranked by $\overline{\mathrm{RMSE}}_{\mathrm{CV5}}$
\State \textbf{return} \texttt{Chaos} if $R^2 > 0.5$, else \texttt{Noise}
\end{algorithmic}
\normalsize
\end{boxedalgorithm}
\end{widetext}

\begin{acknowledgments}
A. L. Chanu and J.-M. Park were supported by an appointment to the JRG Program at the APCTP through the Science and Technology Promotion Fund and Lottery Fund of the Korean Government, also by the Korean Local Governments - Gyeongsangbuk-do Province and Pohang City.
This work is also supported by the National Research Foundation (NRF) of Korea grant funded by the Korea government (MSIT) (RS-2025-00557038) (A. L. Chanu and J.-M. Park).
J. Choi was supported by a KIAS Individual Grant (No. AP092903) via the Center for AI and Natural Sciences at the Korea Institute for Advanced Study (KIAS). This work is supported by the Center for Advanced Computation at KIAS.
A. L. Chanu was also partially supported by the National Research Foundation (NRF) of Korea under grant No. RS-2024-00343900.
This work benefited from discussions during the program ``1st APCTP-SISSA Joint Workshop on AI and Theoretical Physics'' in December 2024 supported by APCTP.
We also express our sincere gratitude to Dr.\ Vandertone Santos Machado (Federal University of Paran\'{a}) for kindly providing the Chua's circuit experiment (ChuaExp) dataset.
\end{acknowledgments}

\appendix

\section{Synthetic and empirical datasets}
\label{app:datasets}

\subsection{Synthetic chaotic systems}
\label{app:chaos}

All synthetic time series used in our analysis have a total length of $L_\mathrm{given} = 10{,}000$ samples. The synthetic chaotic datasets comprise both discrete-time and continuous-time systems. Discrete-time systems are iterated directly. Continuous-time systems are integrated numerically using a fourth-order Runge--Kutta method with a fixed integration time step of $\Delta t = 0.01$. The synthetic chaotic systems used in this work are listed below, together with their governing equations, parameter values, and brief descriptions.

For each synthetic system, a single scalar observable is used as the reservoir input. Specifically, we use the scalar map variable $x_n$ for LogMap and SkTent, the juvenile density $J_n$ for TwoAge, the first state variable $x_1$ for the Lorenz system, the R\"ossler system, and Chua's circuit, and the nuclear NF-$\kappa$B concentration $N_n$ for NFKB. After sampling, each of these scalar time series is denoted uniformly by $x_t$ throughout the main text.

\begin{itemize}

\item \textbf{Logistic map (LogMap):}
\begin{equation}
x_{n+1} = r\,x_n(1 - x_n).
\end{equation}
We use $r = 3.8$, and the initial value $x_0 = 0.1$.

\item \textbf{Skew tent map (SkTent):}
\begin{equation}
x_{n+1} =
\begin{cases}
x_n / a, & x_n < a, \\[4pt]
(1 - x_n)/(1 - a), & x_n \geq a.
\end{cases}
\end{equation}
We use $a = 0.4$, and $x_0 = 0.2$.

\item \textbf{Density-dependent two-age-classes model (TwoAge):}
\begin{equation}
\begin{aligned}
J_{n+1} &= r\,A_n \exp\!\bigl[-\alpha(A_n + J_n)\bigr], \\
A_{n+1} &= c_J\,J_n \exp\!\bigl[-\beta_J(A_n + J_n)\bigr] \\
         &\quad+ c_A\,A_n \exp\!\bigl[-\beta_A(A_n + wJ_n)\bigr],
\end{aligned}
\end{equation}
where $J_n$ and $A_n$ denote the juvenile and adult population densities. We use the values of the parameters $(r,\,\alpha,\,c_J,\,\beta_J,\,c_A,\,\beta_A,\,w) = (118,\,0.001,\,0.2,\,0.07,\,0.8,\,0.05,\,0.5)$, and $(J_0, A_0) = (10, 1)$.

The two-age-classes model describes a density-dependent ecological system in which the strange attractor is split into several disconnected domains~\cite{cazelles1992predictable}. This fragmentation produces a strong overall periodicity that tends to mask the underlying chaotic dynamics, making it a rigorous test case for chaos-detection algorithms~\cite{barahona1996detection}.

\item \textbf{Lorenz system:}
\begin{equation}
\begin{aligned}
\dot{x}_1 &= \sigma(x_2 - x_1), \\
\dot{x}_2 &= x_1(\rho - x_3) - x_2, \\
\dot{x}_3 &= x_1 x_2 - \beta x_3.
\end{aligned}
\end{equation}
We take $\sigma = 10$, $\rho = 28$, $\beta = 8/3$, and the initial condition $(x_1, x_2, x_3)_0 = (1, 1, 1)$, with a sampling ratio of 10 ($\Delta t_{\mathrm{sample}} = 0.1$).

\item \textbf{R\"{o}ssler system:}
\begin{equation}
\begin{aligned}
\dot{x}_1 &= -x_2 - x_3, \\
\dot{x}_2 &= x_1 + a\,x_2, \\
\dot{x}_3 &= b + x_3(x_1 - c).
\end{aligned}
\end{equation}
We use $a = 0.2$, $b = 0.2$, $c = 5.7$, and $(x_1, x_2, x_3)_0 = (0, 1, 1)$, with a sampling ratio of 10 ($\Delta t_{\mathrm{sample}} = 0.1$).

\item \textbf{Chua's circuit:}
\begin{equation}
\begin{aligned}
\dot{x}_1 &= \alpha\bigl(x_2 - x_1 - h(x_1)\bigr), \\
\dot{x}_2 &= x_1 - x_2 + x_3, \\
\dot{x}_3 &= -\beta\,x_2,
\end{aligned}
\end{equation}
where $h(x)$ is the voltage-current characteristic of the Chua diode, given by the piecewise-linear function
\begin{equation}
h(x) = m_1 x + \tfrac{1}{2}(m_0 - m_1)\bigl(|x+1| - |x-1|\bigr).
\end{equation}
We use $\alpha = 15.6$, $\beta = 28$, $m_0 = -1.143$, $m_1 = -0.714$, and $(x_1, x_2, x_3)_0 = (0.1, 0, 0)$, with a sampling ratio of 10 ($\Delta t_{\mathrm{sample}} = 0.1$).

\item \textbf{NF-$\kappa$B gene regulatory network model (NFKB):}
\begin{align}
\dot{N}_n &= k_{Nin}(N_{\mathrm{tot}} - N_n)\frac{K_I}{K_I + I}
            - k_{Iin}\,I\,\frac{N_n}{K_N + N_n}, \\[4pt]
\dot{I}_{\mathrm{RNA}} &= k_t\,N_n^{2} - \gamma_m\,I_{\mathrm{RNA}}, \\[4pt]
\dot{I} &= k_{tl}\,I_{\mathrm{RNA}}
          - \alpha\,IKK_a(N_{\mathrm{tot}} - N_n)\frac{I}{K_I + I}, \\[4pt]
\dot{IKK}_a &= k_a \cdot TNF \cdot IKK_n - k_i\,IKK_a, \\[4pt]
\dot{IKK}_i &= k_i\,IKK_a
              - k_p\,IKK_i\,\frac{k_{A20}}{k_{A20} + [A20]\cdot TNF}, \\[4pt]
IKK_n &= [IKK]_{\mathrm{tot}} - IKK_a - IKK_i, \\[4pt]
TNF &= 0.5 + A\sin\!\left(\frac{2\pi}{T}\,t\right).
\end{align}
Here, $N_n$ is the nuclear NF-$\kappa$B concentration; $I_{\mathrm{RNA}}$ is the I$\kappa$B mRNA level; $I$ is the cytoplasmic I$\kappa$B protein concentration; $IKK_a$ and $IKK_i$ are the active and inactive forms of I$\kappa$B kinase; and $TNF$ is a sinusoidal external drive with amplitude $A$ and period $T$. Values of the model parameters are taken from Supplementary Table~I in Ref.~\cite{heltberg2019chaotic}.

NF-$\kappa$B is a transcription factor that regulates numerous genes involved in immune responses and has been widely studied in cancer research. We consider the five-dimensional gene regulatory network model introduced in Ref.~\cite{heltberg2019chaotic}, in which periodic stimulation by a cytokine-like signal called tumor necrosis factor (TNF) can drive the NF-$\kappa$B dynamics into a chaotic regime.

\end{itemize}

\subsection{Synthetic noise processes}
\label{app:stochastic}

We generate time series from the following noise processes with total length $L_{\mathrm{given}} = 10{,}000$ samples.

\begin{itemize}

\item \textbf{Gaussian noise (Gauss):}
An i.i.d.\ sequence drawn from the standard normal distribution, $X_t \sim \mathcal{N}(0, 1)$.

\item \textbf{Uniform noise (Uniform):}
An i.i.d.\ sequence drawn from a uniform distribution on $(0,1)$, $X_t \sim \mathcal{U}(0, 1)$ with a flat probability density.

\item \textbf{Cauchy noise (Cauchy):}
An i.i.d.\ sequence drawn from a Cauchy distribution generated as $X_t = \tan\!\bigl(\pi(u_t - 0.5)\bigr)$, where $u_t \sim \mathcal{U}(0,1)$.

\item \textbf{Flicker noise (Flicker):}
A stationary noise process with power spectral density $S(f) \propto 1/f^{\alpha}$, where $\alpha = 1$. We generate the sequence using the method of Timmer and K\"{o}nig~\cite{timmer1995}.

\item \textbf{Fractional Gaussian noise (fGn):}
A stationary Gaussian process with long-range, persistent temporal
correlations. We generate the sequence using Hosking's sequential
algorithm~\cite{hosking1984} with Hurst exponent $H = 0.7$.

\item \textbf{Fractional Brownian motion (fBm):}
A non-stationary, self-similar Gaussian process with Hurst exponent
$H = 0.7$, obtained as the cumulative sum of the fractional Gaussian
noise described above.

\end{itemize}

\subsection{Empirical datasets}
\label{app:emp_datasets}

We evaluate the proposed method on a collection of empirical time series data spanning biomedical, physical, and geophysical systems. Unless otherwise stated below, the empirical datasets are analyzed at their full available lengths, with the normalization in Eq.~\eqref{eq:preprocess} as the only preprocessing step. The specific time series length $L_\mathrm{given}$ for each empirical dataset is listed in Table~\ref{tab:hp_empirical}.

\begin{itemize}

\item \textbf{Voice recording (VOICE):}
We use a single voice-sample record from the VOICED (VOice ICar fEDerico II) Database v1.0.0, provided as WFDB files (\texttt{voice001.hea} and \texttt{voice001.dat}) sampled at 8\,kHz (38,080 samples). To focus on sustained phonation, the first $2{,}000$ samples (${\approx}\,0.25$\,s at 8\,kHz) containing initial silence and voice onset are discarded, leaving 36,080 samples for analysis. The recording is labeled as a sample from a 32-year-old male with a clinical diagnosis of hyperkinetic dysphonia and is accompanied by metadata such as VHI/RSI scores and lifestyle details~\cite{CESARI2018310}. Previous studies have identified \textit{chaotic} dynamics in voice production systems, including evidence of nonlinear and chaotic oscillations in the vocal fold~\cite{herzel1995, tao2008chaotic}.

\item \textbf{Laser intensity pulsing (LaserP):}
We use the intensity pulsing data of a far-infrared laser from the Santa Fe Time Series Competition~\cite{SantaFeA2}, which have previously been identified as exhibiting \textit{chaotic} dynamics~\cite{kulp2014discriminating, hubner1989dimensions}.

\item \textbf{Squid giant axon membrane potential (SGAMP):}
We use the membrane-potential recording from the Squid Giant Axon database, scaled to millivolts (mV) to capture the axon's physiological response~\cite{paydarfar2006noisy}. Previous studies have reported \textit{chaotic} oscillations and bifurcations in squid giant axons~\cite{aihara1987, mees1992}. We also include a 10:1 downsampled version (SGAMP10) to separate the effect of sampling rate from time series length.

\item \textbf{Chua's circuit experiment (ChuaExp):}
We use experimental voltage recordings measured across two capacitors ($C_1$ and $C_2$) of a physical Chua's circuit implementation, obtained directly from the authors of Ref.~\cite{prado2020parameter}.
The recorded voltages are quantized in increments of $0.2$\,V and take on only 69 distinct values, resulting in limited measurement resolution.
The dynamics are consistent with \textit{chaotic} behavior, in agreement with the well-established chaotic regime of Chua's circuit~\cite{chua1994, prado2020parameter}.

\item \textbf{Parkinsonian tremor (ParkTr):}
From Ref.~\cite{JETI1937_1_1}, we use the Parkinsonian tremor time series \texttt{1937-1\_1}, originally analyzed by Timmer et al.~\cite{timmer2000pathological}. The nature of this signal has been the subject of a long-standing debate. Through systematic nonlinear time series analysis, Timmer et al.~\cite{timmer2000pathological} concluded that the dynamics lack the defining features of deterministic chaos and are instead consistent with a \textit{nonlinear stochastic oscillator}, a view further supported by characterizations of the tremor as a diffusional process~\cite{gao2002pathological}. Other studies, however, report that Parkinsonian tremor exhibits chaotic dynamics~\cite{gantert1992analyzing,sadeghirazlighi2012study}, and a recent analysis finds both stochastic and chaotic dynamics depending on the severity of the disease, with untreated rest tremor showing chaotic dynamics that become more stochastic under treatment~\cite{sarbaz2020exploring}. Its dynamical classification thus remains unsettled.

\item \textbf{North Atlantic Oscillation index (NAOidx):}
We use the daily mean NAO index from the NOAA Climate Prediction Center~\cite{NAO_CPC}.
We analyze $L_\mathrm{given}=27{,}676$ daily samples starting January 1950. A previous study reported a stochastic origin for these data~\cite{toker2020simple}.

\item \textbf{Sunspot number (Sunspot):}
We use the daily total sunspot number (V2.0) from the Sunspot Index and Long-term Solar Observations (SILSO) repository~\cite{sp}. To avoid missing values in the early records, we use the most recent $20{,}000$ days (from 1970-12-29 to 2025-09-30). A previous analysis suggested a stochastic origin for these data~\cite{boaretto2021discriminating}.

\item \textbf{Heart rate RR-interval (RR):}
We use five RR-interval recordings from each of three subject groups comprising healthy subjects, patients with congestive heart failure, and patients with atrial fibrillation~\cite{glass2009introduction, PhysioNetChaosHR}. The original RR-interval recordings contain substantially more than $10{,}000$ intervals. For the present analysis, we use $10{,}000$ consecutive intervals, corresponding to samples $20{,}001$--$30{,}000$ of each recording, to ensure a common data length across subjects. We denote recordings from healthy subjects as \textbf{RRnorm}, from patients with congestive heart failure as \textbf{RRchf}, and from patients with atrial fibrillation as \textbf{RRaf}. Previous studies have suggested a stochastic origin for these data~\cite{toker2020simple, boaretto2021discriminating}.

\end{itemize}

\section{Optimized hyperparameters}
\label{app:hp}

Tables~\ref{tab:hp_chaotic} and~\ref{tab:hp_stochastic} list the optimized hyperparameter values for the synthetic chaotic systems and noise processes, respectively, under the value-to-difference cross-prediction scheme. The six hyperparameters are the reservoir size $n$, spectral radius $\rho$, input scaling $\sigma_{\mathrm{in}}$, leak rate $a_{\mathrm{leak}}$, ridge regularization coefficient $\beta$, and sparsity parameter $k$. We fix the time series length $L_\mathrm{given} = 10{,}000$ for all synthetic systems. Table~\ref{tab:hp_empirical} lists the corresponding values for the empirical datasets ($L_\mathrm{given}$ varies by system).

\begin{table*}
\centering
\caption{Optimized hyperparameters for synthetic chaos ($L_\mathrm{given} = 10{,}000$).}
\label{tab:hp_chaotic}
\renewcommand{\arraystretch}{1.05}
\setlength{\tabcolsep}{6pt}
\begin{tabular}{lcccccc}
\hline\hline
System   & $n$ & $\rho$ & $\sigma_{\mathrm{in}}$ & $a_{\mathrm{leak}}$ & $\beta$ & $k$ \\
\hline
LogMap      & 455 & 0.342 & 1.552 & 0.862 & $6.53\times10^{-7}$ & 0.151 \\
SkTent      & 492 & 0.434 & 2.708 & 0.973 & $6.64\times10^{-5}$ & 0.865 \\
TwoAge      & 667 & 1.647 & 2.535 & 0.900 & $3.75\times10^{-5}$ & 0.221 \\
Chua        & 402 & 0.701 & 2.813 & 0.819 & $1.41\times10^{-4}$ & 0.779 \\
Lorenz      & 638 & 0.938 & 3.000 & 0.785 & $5.74\times10^{-4}$ & 0.604 \\
Rossler & 510 & 0.939 & 2.634 & 0.848 & $1.55\times10^{-6}$ & 0.879 \\
NFKB        & 511 & 0.937 & 2.634 & 0.849 & $2.99\times10^{-6}$ & 0.879 \\
\hline\hline
\end{tabular}
\end{table*}

\begin{table*}
\centering
\caption{Optimized hyperparameters for synthetic noise ($L_\mathrm{given} = 10{,}000$).}
\label{tab:hp_stochastic}
\renewcommand{\arraystretch}{1.05}
\setlength{\tabcolsep}{6pt}
\begin{tabular}{lcccccc}
\hline\hline
System   & $n$ & $\rho$ & $\sigma_{\mathrm{in}}$ & $a_{\mathrm{leak}}$ & $\beta$ & $k$ \\
\hline
Gauss & 603 & 0.100 & 2.885 & 0.238 & $6.66\times10^{-4}$ & 0.773 \\
Uniform  & 190 & 0.657 & 0.821 & 0.255 & $2.05\times10^{-4}$ & 0.691 \\
Flicker  & 232 & 0.896 & 0.033 & 0.561 & $3.88\times10^{-4}$ & 0.939 \\
Cauchy   & 75  & 0.226 & 0.010 & 0.011 & $7.99\times10^{-5}$ & 0.010 \\
fBm      & 70  & 0.881 & 0.065 & 0.537 & $3.56\times10^{-4}$ & 0.933 \\
fGn      & 133 & 0.836 & 0.053 & 0.548 & $4.29\times10^{-4}$ & 0.892 \\
\hline\hline
\end{tabular}
\end{table*}

\begin{table*}
\centering
\caption{Optimized hyperparameters for empirical datasets. The upper block lists datasets in the reference chaos category, and the lower block lists those in the reference noise category. The time series length $L_\mathrm{given}$ is given for each system.}
\label{tab:hp_empirical}
\renewcommand{\arraystretch}{1.05}
\setlength{\tabcolsep}{5pt}
\begin{tabular}{lccccccc}
\hline\hline
System & $L_\mathrm{given}$ & $n$ & $\rho$ & $\sigma_{\mathrm{in}}$ & $a_{\mathrm{leak}}$ & $\beta$ & $k$ \\
\hline
\multicolumn{8}{l}{\textit{Reference chaos category}} \\
\hline
VOICE    & 36{,}080  & 793 & 1.017 & 0.662 & 0.519 & $1.43\times10^{-4}$ & 0.371 \\
LaserP   & 9{,}093   & 553 & 0.989 & 2.530 & 0.721 & $5.30\times10^{-4}$ & 0.597 \\
SGAMP    & 125{,}000 & 696 & 1.005 & 1.669 & 0.992 & $1.17\times10^{-4}$ & 0.126 \\
SGAMP10  & 12{,}500  & 516 & 0.935 & 2.691 & 0.865 & $1.03\times10^{-8}$ & 0.886 \\
ChuaExp  & 20{,}000  & 967 & 0.647 & 2.079 & 0.930 & $4.57\times10^{-4}$ & 0.557 \\
ParkTr   & 30{,}000  & 130 & 1.139 & 0.110 & 0.608 & $1.83\times10^{-4}$ & 0.746 \\
\hline
\multicolumn{8}{l}{\textit{Reference noise category}} \\
\hline
NAOidx   & 27{,}676 & 353 & 0.617 & 0.588 & 0.923 & $8.51\times10^{-4}$ & 0.800 \\
Sunspot  & 20{,}000 & 227 & 0.982 & 0.103 & 0.382 & $2.34\times10^{-4}$ & 0.857 \\
RRnorm-1 & 10{,}000 & 635 & 0.411 & 2.976 & 0.435 & $4.29\times10^{-4}$ & 0.064 \\
RRnorm-2 & 10{,}000 & 136 & 0.737 & 0.264 & 0.247 & $2.36\times10^{-4}$ & 0.507 \\
RRnorm-3 & 10{,}000 & 325 & 0.604 & 1.408 & 0.838 & $6.97\times10^{-4}$ & 0.840 \\
RRnorm-4 & 10{,}000 & 52  & 1.362 & 0.136 & 0.533 & $1.60\times10^{-4}$ & 0.791 \\
RRnorm-5 & 10{,}000 & 58  & 0.959 & 0.126 & 0.480 & $4.25\times10^{-4}$ & 0.964 \\
RRchf-1  & 10{,}000 & 87  & 0.231 & 2.048 & 1.000 & $4.18\times10^{-4}$ & 0.322 \\
RRchf-2  & 10{,}000 & 58  & 0.919 & 2.862 & 1.000 & $1.57\times10^{-4}$ & 0.698 \\
RRchf-3  & 10{,}000 & 266 & 1.145 & 0.199 & 0.569 & $1.83\times10^{-4}$ & 0.689 \\
RRchf-4  & 10{,}000 & 189 & 1.184 & 0.089 & 0.614 & $2.09\times10^{-4}$ & 0.768 \\
RRchf-5  & 10{,}000 & 514 & 0.819 & 0.665 & 0.942 & $9.03\times10^{-4}$ & 0.781 \\
RRaf-1   & 10{,}000 & 55  & 0.896 & 0.446 & 0.227 & $2.19\times10^{-4}$ & 0.504 \\
RRaf-2   & 10{,}000 & 64  & 1.621 & 0.337 & 0.199 & $3.70\times10^{-4}$ & 0.366 \\
RRaf-3   & 10{,}000 & 87  & 0.115 & 0.800 & 0.141 & $5.16\times10^{-5}$ & 0.927 \\
RRaf-4   & 10{,}000 & 55  & 0.193 & 0.510 & 0.995 & $9.96\times10^{-4}$ & 0.810 \\
RRaf-5   & 10{,}000 & 51  & 0.100 & 2.509 & 0.445 & $1.62\times10^{-4}$ & 0.679 \\
\hline\hline
\end{tabular}
\end{table*}

\FloatBarrier

\bibliographystyle{apsrev4-2}
\bibliography{paper}

\end{document}